\documentclass[12pt]{article}

\pdfoutput=1
\usepackage[latin9]{inputenc}
\usepackage[top=80pt,bottom=85pt,left=67pt,right=67pt]{geometry}
\usepackage{amssymb,amsmath}
\usepackage{xypic,subfigure}
\usepackage{graphicx,color}
\usepackage{float}
\usepackage{cite}
\usepackage[debug,pageanchor=false]{hyperref}
\definecolor{link}{rgb}{.8,.15,.1}
\hypersetup{colorlinks=true,linkcolor=link,citecolor=link,urlcolor=link,linktocpage}

\makeatletter
\@addtoreset{equation}{section}
\makeatother

\renewcommand{\theequation}{\thesection.\arabic{equation}}

\newcommand{\beq}{\begin{equation}}
\newcommand{\eeq}{\end{equation}}
\newcommand{\bea}{\begin{eqnarray}}
\newcommand{\eea}{\end{eqnarray}}

\newcommand{\eq}{\begin{equation}}
\newcommand{\feq}{\end{equation}}
\newcommand{\eqn}{\begin{eqnarray}}
\newcommand{\feqn}{\end{eqnarray}}

\newcommand{\mrm}[1]{\mbox{$\mathrm{#1}$}}

\begin{document}
\begin{titlepage}

\begin{center}

\vskip .5in 
\noindent

{\Large \bf{Line defects as brane boxes in Gaiotto-Maldacena geometries}}

\bigskip\medskip

Yolanda Lozano$^{a,b}$\footnote{ylozano@uniovi.es},  Nicol\`o Petri$^c$\footnote{petri@post.bgu.ac.il}, Cristian Risco$^{a,b}$\footnote{cristianrg96@gmail.com}  \\

\bigskip\medskip
{\small 

a: Department of Physics, University of Oviedo,
Avda. Federico Garcia Lorca s/n, 33007 Oviedo}

\medskip
{\small and}

\medskip
{\small 

b: Instituto Universitario de Ciencias y Tecnolog\'ias Espaciales de Asturias (ICTEA),\\
Calle de la Independencia 13, 33004 Oviedo, Spain}

\bigskip\medskip
{\small 

c: Department of Physics, Ben-Gurion University of the Negev, Be'er Sheva 84105, Israel}

\vskip 2cm 

     	{\bf Abstract }
     	\end{center}
     	\noindent
	
We construct a new family of AdS$_2\times S^2\times S^2$ solutions to Type IIA supergravity with 4 supercharges acting with non-Abelian T-duality on the recent class constructed in \cite{Lozano:2021fkk}. We focus on a particular solution in this class asymptoting locally to an AdS$_5$ Gaiotto-Maldacena geometry. This solution allows for a line defect interpretation within the 4d $\mathcal{N}=2$ SCFT dual to this geometry, that we study in detail. We show that the defect branes, consisting on a non-trivial intersection of D2-D4-NS5-F1 branes, can be interpreted as baryon vertices within the 4d $\mathcal{N}=2$ SCFT, whose backreaction gives rise to the AdS$_2$ solution. We construct the explicit quiver quantum mechanics that flows in the IR to the dual SCQM, and show that it can be embedded within the quiver CFT associated to the AdS$_5$ solution. The quiver quantum mechanics arises from a brane box set-up of D2-branes stretched between perpendicular NS5-branes, that we construct from the AdS$_2$ solution. We provide non-trivial checks of our proposed duality. Our construction provides one further example of the successful applications of non-Abelian T-duality to holography, in this case in providing a very non-trivial connection between AdS$_2$ solutions, line defects and brane boxes.
    	
\noindent

\vfill
\eject

\end{titlepage}

\setcounter{footnote}{0}

\tableofcontents

\setcounter{footnote}{0}
\renewcommand{\theequation}{{\rm\thesection.\arabic{equation}}}

\section{Introduction}

As it is well-known, the study of the AdS$_2$/CFT$_1$ correspondence is of paramount importance to understanding the microscopical description of extremal black holes. Notable efforts have been devoted to this study in the last decades. More recently it has also been shown to play a prominent role in the description of conformal line defects. These are one dimensional defects that preserve a superconformal subalgebra of the superconformal algebra of the field theory  where they are embedded. Being conformal holography provides a very powerful tool for their study \cite{Karch:2000gx,DeWolfe:2001pq, Aharony:2003qf, DHoker:2006vfr,Lunin:2007ab}.

Defect conformal field theories are typically engineered in terms of a brane intersection consisting on {\it defect branes} ending on a bound state of {\it background branes}, in which a higher dimensional CFT lives. Holographically this is described by low dimensional AdS spaces with non-compact internal manifolds, that reproduce a higher dimensional AdS geometry asymptotically. The presence of the non-compact direction renders the defect field theory ill-defined, but this can be interpreted as the need to complete this CFT by the higher dimensional one far away from the defects. Holographically the associated divergence of the central charge (or free energy) is absorbed when the non-compact coordinate becomes part of the higher dimensional AdS space. This approach to the study of defects has been very fruitful, with substantial amount of work being devoted to identifying low dimensional AdS solutions dual to defect CFTs in recent years  \cite{DHoker:2007mci,Chiodaroli:2009yw,Chiodaroli:2009xh,Dibitetto:2017tve,Dibitetto:2017klx,Gutperle:2017nwo,Dibitetto:2018iar,Dibitetto:2018gtk,Gutperle:2018fea,Chen:2019qib,Gutperle:2019dqf,Chen:2020mtv,Faedo:2020nol,Faedo:2020lyw,Chen:2020efh,Chen:2021mtn,Lozano:2021fkk,Gutperle:2022pgw,Gutperle:2022fma,Lozano:2022ouq, Lozano:2022swp}.
Interestingly, in some of these realisations it has been possible to embed the defect CFT within the higher dimensional theory through explicit quiver-like constructions \cite{Faedo:2020nol,Faedo:2020lyw,Lozano:2021fkk,Lozano:2022ouq}. In this work we will provide one further example of AdS$_2$ background with a line defect interpretation, and we will construct an explicit one dimensional quiver describing the line defect CFT.

The study of the AdS$_2$/CFT$_1$ correspondence is well-known to offer unique challenges compared to its higher dimensional counterparts. These have to do mainly with the non-connectedness of the boundary of AdS$_2$, where the putative dual quantum mechanics lives, and with the peculiarity of AdS$_2$ gravity, which does not support finite energy excitations \cite{Maldacena:1998uz,Strominger:1998yg,Hartman:2008dq,Alishahiha:2008tv}. Interesting ways to avoid these difficulties have been proposed in the literature \cite{Balasubramanian:2003kq,Balasubramanian:2009bg,Almheiri:2014cka,Maldacena:2016hyu,Maldacena:2016upp,Harlow:2018tqv,Dibitetto:2019nyz,Aniceto:2020saj,Lozano:2020txg,Lozano:2020sae,Lozano:2021rmk,Ramirez:2021tkd,Lozano:2021fkk}. These range from the study of nearly AdS$_2$ nearly CFT$_1$ dual pairs to the construction of AdS$_2$ solutions as null orbifolds of AdS$_3$ backgrounds, for which the 1d superconformal quantum mechanics (SCQM) is realised as a discrete light-cone compactification of a 2d CFT \cite{Strominger:1998yg,Balasubramanian:2003kq,Balasubramanian:2009bg,Aniceto:2020saj,Lozano:2020txg}. Our results suggest that the SCQM that we will construct in this paper belongs to this second class. 

Our work is organised as follows. In section 2 we start presenting a new class of AdS$_2$ solutions to Type IIA supergravity preserving $\mathcal{N}=4$ supersymmetries. We construct these solutions by performing a non-Abelian T-duality transformation (with respect to a freely acting SU(2) compact group) on the class of solutions to Type IIB supergravity recently constructed in \cite{Lozano:2021fkk}. These solutions describe D1-F1-D5-NS5 branes ending on D3-branes and preserve $\mathcal{N}=4$ supersymmetry (small). For a given brane profile they asymptote locally to AdS$_5\times S^5$, and should thus find an interpretation as dual to line defect CFTs. In \cite{Lozano:2021fkk} some evidence was gathered that led us to propose that the line defect operators these solutions are dual to are baryon vertices in $\mathcal{N}=4$ SYM. We show that after the duality these solutions describe D2-F1-D4'-NS5' branes ending on D4-D6-NS5 branes, and for a particular brane profile asymptote locally to the Gaiotto-Maldacena geometry \cite{Gaiotto:2009gz} constructed in \cite{Sfetsos:2010uq}, that arises by acting with non-Abelian T-duality on AdS$_5\times S^5$. The resulting solution should then allow for a line defect interpretation. In section 3 we move on to the study of the defect superconformal quantum mechanics. We start discussing the brane set-up associated to the solution, given by a D2-brane box model \cite{Hanany:1997tb,Hanany:1998it,Hanany:2018hlz}, in which D2 colour branes are stretched between NS5 and NS5' branes in two perpendicular directions. We analyse the quantisation of the open strings stretched between the D2-branes in the different boxes, as well as the role played by the D4-D4' and D6 branes of the brane intersection. From here we propose a planar quiver QM which can be interpreted as embedded in the linear quiver that describes the 4d $\mathcal{N}=2$ SCFT \cite{Witten:1997sc} where the defects live. This SCFT is the one dual to the Gaiotto-Maldacena geometry to which the AdS$_2$ solution flows asymptotically locally in the UV, studied in \cite{Lozano:2016kum}. Our SCQMs are far more elaborated than those proposed so far in the literature, including the ones discussed in \cite{Lozano:2021fkk,Lozano:2020txg,Lozano:2020sae,Lozano:2021rmk,Ramirez:2021tkd, Assel:2018rcw,Assel:2019iae}, which consist on linear quivers. 
We propose a way of calculating the central charge of the SCQM that we show agrees with the holographic calculation in the holographic limit, and discuss the reasons behind this agreement. Finally, we turn to discuss the massive F1-strings present in the brane intersection. We show that they are interpreted, together with the D4' and the D2 branes, as baryon vertices in the 4d SCFT. This allows us to interpret the AdS$_2$ solution as the backreacted geometry that arises after the baryon vertices are introduced in the 4d SCFT. In section 4 we present our conclusions and open problems. We emphasise the interesting role played, once more, by non-Abelian T-duality in the context of holography, in this case in allowing to construct holographic defects  described by an elaborated brane box model. We emphasise as well  that this is to our knowledge the first time that a brane box construction is realised holographically, other than the specific circular brane boxes discussed in \cite{Faedo:2020lyw}, dual to the  AdS$_3$ solutions constructed therein. We include details of the non-Abelian T-duality transformation that we have used to construct our new class of solutions in Appendix \ref{first}. In Appendix \ref{second} we present a black hole geometry constructed through non-Abelian T-duality acting on the brane intersection that underlies the AdS$_2$ solutions in Type IIB that have been the starting point of our analysis in this paper. The resulting solution gives rise to the new class of solutions discussed in section \ref{Newclass} in the near horizon limit, but presents some obstacles towards a possible interpretation as a brane intersection. This is a typical situation for AdS solutions constructed through non-Abelian T-duality, whose underlying brane set-up is not known even in the simplest examples (see \cite{Terrisse:2018hhf}). The interest of the background that we present in Appendix \ref{second} is that it describes a 4d black hole with $\mathcal{N}=4$ supersymmetry that should be interesting to explore in the context of black hole physics.

\section{A new class of AdS$_2\times S^2\times S^2$ solutions to Type IIA via non-Abelian T-duality}
\label{Newclass}

In this section we take as our starting point the class of AdS$_2\times S^3\times S^2$ solutions to Type IIB supergravity recently constructed in \cite{Lozano:2021fkk}\footnote{In  \cite{Lozano:2021fkk} the more general case in which the $S^5$ is orbifolded by $\mathbb{Z}_k$ was considered. Here we will restrict to the case $k=1$. This does not affect the number of preserved supersymmetries.}. Besides providing a new class of AdS$_2$ geometries with $\mathcal{N}=4$ supersymmetries these solutions had the interesting property of including a particular background that asymptotes locally to AdS$_5\times S^5$ in the UV, and thus allows for an interpretation as holographic dual to a line defect CFT embedded in 4d $\mathcal{N}=4$ SYM. In \cite{Lozano:2021fkk} some evidence was gathered that led us to propose that the line defect operators these solutions are dual to are baryon vertices in $\mathcal{N}=4$ SYM. These operators would be realised in string theory as $(p,q)$ strings stretched between stacks of $(q,p)$ 5-branes and D3 colour branes, generalising the constructions in \cite{Yamaguchi:2006tq,Gomis:2006sb} by acting with SL(2,$\mathbb{Z}$).  
It was also shown in \cite{Lozano:2021fkk} that 
 these solutions are mapped through Abelian T-duality to AdS$_2\times S^2\times S^2$ solutions to Type IIA supergravity arising in the near horizon limit of F1-D2-D4'-NS5' defect branes 
 embedded in the Type IIA realisation of 4d $\mathcal{N}=4$ SYM, namely the semi-localised D4-NS5 brane intersection studied in \cite{Youm:1999ti,Fayyazuddin:1999zu,Loewy:1999mn,Alishahiha:1999ds,Oz:1999qd,Lozano:2016kum}. In turn, this intersection gives rise in the near horizon limit to a Gaiotto-Maldacena geometry, to which the IIA AdS$_2$ solutions flow asymptotically in the UV. 

In this section we construct more general AdS$_2$ geometries in Type IIA supergravity allowing for a similar holographic interpretation as line defects within 4d $\mathcal{N}=2$ SCFTs dual to Gaiotto-Maldacena geometries. The way we construct these geometries is by acting with non-Abelian T-duality on the class of  AdS$_2\times S^3\times S^2$ solutions to Type IIB supergravity constructed in \cite{Lozano:2021fkk}. The new backgrounds are fibrations of AdS$_2\times S^2\times S^2$ over four intervals and contain a particular solution that asymptotes locally to the Gaiotto-Maldacena geometry constructed in \cite{Sfetsos:2010uq}. We will show that it is possible to give an explicit interpretation for this AdS$_2$ solution as a baryon vertex defect embedded in the 4d $\mathcal{N}=2$ SCFT dual to the Gaiotto-Maldacena geometry, studied in \cite{Lozano:2016kum}.


Our starting point is the class of solutions given by equation (3.13) in \cite{Lozano:2021fkk}, where we do not include KK-monopoles. The solutions are given by
\begin{equation}\label{solutionsIIB}
	\begin{aligned}
		ds_{10}^2 &= q_{\text{D}1}^{3/2}q_{\text{F}1}^{1/2}    H_{\text{D}3}^{-1/2}\left(ds^2_{\text{AdS}_2}+ds^2_{S^2}\right) +q_{\text{D}1}^{1/2}q_{\text{F}1}^{-1/2} H_{\text{D}3}^{1/2} \left(dy^2+dz^2+ dr^2 + r^2 ds^2_{S^3}\right)\,,\\
		e^{\Phi}&=q_{\text{D}1}q_{\text{F}1}^{-1}   \,,\qquad H_{3} = -q_{\text{D}1}\text{vol}_{\text{AdS}_2}\wedge dz-q_{\text{D}1} \text{vol}_{  S^2}\wedge dy\,,\\
		F_{3}&=-q_{\text{F}1}\text{vol}_{\text{AdS}_2}\wedge dy+q_{\text{F}1} \text{vol}_{  S^2}\wedge dz\,,\\
		F_{5}&= d[q_{\text{D}1}^{2}q_{\text{F}1}^{2}  H_{\text{D}3}^{-1}\,\text{vol}_{\text{AdS}_2}\wedge \text{vol}_{ S^2}]+\\
		&+r^3[
		(\partial_z H_{\text{D}3} dy-\partial_y H_{\text{D}3} dz) \wedge  dr -\partial_r H_{\text{D}3} dy \wedge dz] \wedge \text{vol}_{ S^3}\,,
	\end{aligned}
\end{equation}
with $H_{\text{D} 3}$ satisfying the master equation
\begin{equation}\label{mastereq}
\nabla^2_{\mathbb{R}^4_r}H_{\text{D}3}+\nabla^2_{\mathbb{R}^2_{(y,z)}}H_{\text{D}3}=0.
\end{equation}
These solutions arise in the near horizon limit of the brane intersection depicted in Table \ref{Table:D1-F1-D5-NS5-D3}, and preserve small $\mathcal{N}=4$ supersymmetry, with the SU(2) R-symmetry group realised on the $S^2$. 
\begin{table}[http!]
\renewcommand{\arraystretch}{1}
\begin{center}
\scalebox{1}[1]{
\begin{tabular}{c| c cc  c c  c  c c c c}
 branes & $t$ & $x^1$ & $x^2$ & $x^3$ & $z$ & $y$ & $x^6$ & $x^7$ & $x^8$ & $x^9$ \\
\hline \hline
$\mrm{D}3$ & $\times$ & $\times$ & $\times$ & $\times$ & $-$ & $-$ & $-$ & $-$ & $-$ & $-$ \\
$\mrm{D}1$ & $\times$ & $-$ & $-$ & $-$ & $-$ & $\times$ & $-$ & $-$ & $-$ & $-$ \\
$\mrm{F}1$ & $\times$ & $-$ & $-$ & $-$ & $\times$ & $-$ & $-$ & $-$ & $-$ & $-$ \\
$\mrm{D}5$ & $\times$ & $-$ & $-$ & $-$ & $-$ & $\times$ & $\times$ & $\times$ & $\times$ & $\times$ \\
$\mrm{NS}5$ & $\times$ & $-$ & $-$ & $-$ & $\times$ & $-$ & $\times$ & $\times$ & $\times$ & $\times$ \\
\end{tabular}
}
\caption{BPS/8 intersection describing D1-F1-D5-NS5 branes ending on D3 branes. $x^1,x^2,x^3$ are the coordinates realising the SO(3) R-symmetry.} \label{Table:D1-F1-D5-NS5-D3}
\end{center}
\end{table}

We now perform a non-Abelian T-duality transformation (the details of which are explained in  Appendix \ref{first}) along the $S^3$, that we parametrise as
\begin{equation}
ds^2_{S^3}=\Bigl(d\psi+\frac{\omega}{2}\Bigr)^2+\frac14 ds^2_{{\tilde S}_2}.
\end{equation}
After this transformation the $S^{3}$ group manifold  is replaced by $\mathbb{R}^3$.
The new class of solutions to Type IIA supergravity reads,
\begin{equation} \label{NATDyz}
	\begin{split} 
		d s_{10}^2 &= q_{\text{D}1}^{3/2}q_{\text{F}1}^{1/2}    H_{\text{D}3}^{-1/2}\left(ds^2_{\text{AdS}_2}+ds^2_{S^2}\right) +q_{\text{D}1}^{1/2}q_{\text{F}1}^{-1/2} H_{\text{D}3}^{1/2}\,  \left( dy^2+dz^2 +dr^2\right)\\
		&+4q_{\text{D}1}^{-1/2}q_{\text{F}1}^{1/2} H_{\text{D}3}^{-1/2} r^{-2} (d\rho^2 + H \rho^2 ds_{\tilde S^2}^2 )\,,\\
		e^{\Phi} &= 8 q_{\text{D}1}^{1/4} q_{\text{F}1}^{-1/4} H_{\text{D}3}^{-3/4} H^{1/2}r^{-3}\,, \\
		B_{2} &= q_{\text{D}1}(z\, \text{vol}_{\text{AdS}_2} + y\, \text{vol}_{S^2})+\frac{16 q_{\text{F}1} \rho^3}{16 q_{\text{F}1} \rho^2 + q_{\text{D}1} H_{\text{D}3} r^4} \, \text{vol}_{\tilde{S}^2}\,,\\
		F_{2} &= -8^{-1} r^3 [(\partial_z H_{\text{D}3} \, dy - \partial_y H_{\text{D}3} \, dz) \wedge dr - \partial_r H_{\text{D}3} \, dy \wedge dz]\,, \\
		F_{4} &=  ( \text{vol}_{\text{AdS}_2} \wedge dy -\text{vol}_{S^2} \wedge dz) \wedge (q_{\text{F}1} \rho \, d\rho - 8^{-1} q_{\text{D}1} H_{\text{D}3} r^3 \,dr)+\\
		&+ \frac{16 q_{\text{F}1} \rho^3}{16 q_{\text{F}1} \rho^2 + q_{\text{D}1} H_{\text{D}3} r^2} \, \text{vol}_{\tilde{S}^2} \wedge F_{2}\,,\\
		F_{6} &= d[q_{\text{D}1}^2 q_{\text{F}1}^2 \rho H_{\text{D}3}^{-1} \, \text{vol}_{\text{AdS}_2} \wedge \text{vol}_{S^2}\wedge d\rho] +\\
		&+ \rho^2 H r^{-2} \,d[q_{\text{F}1} \rho r^2 (\text{vol}_{\text{AdS}_2}  \wedge dy - \text{vol}_{S^2}  \wedge dz) \wedge \text{vol}_{\tilde{S}^2}]\,,
	\end{split}
\end{equation} 
where we have defined
\begin{equation}
H = \frac{q_{\text{D}1} H_{\text{D}3} r^4}{16 q_{\text{F}1} \rho^2 + q_{\text{D}1} H_{\text{D}3} r^4}\,.
\end{equation}
$H_{\text{D}3}$ satisfies \eqref{mastereq} and $(\rho,{\tilde S}^2)$ parametrise the $\mathbb{R}^3$ that arises after the non-Abelian T-duality transformation. Note that as it is common after non-Abelian T-duality the brane intersection from where the AdS geometry arises in the near horizon limit cannot be easily identified\footnote{The reader is referred to \cite{Terrisse:2018hhf} where this is discussed for the AdS$_5$ geometry constructed in \cite{Sfetsos:2010uq}, by performing non-Abelian T-duality on the AdS$_5\times S^5$ background. Even in this simpler example the non-Abelian T-dual of the solution associated to N D3-branes cannot be easily interpreted as a D4-NS5-D6 brane intersection.}. The obvious candidate as brane intersection underlying the solutions \eqref{NATDyz} would be the non-Abelian T-dual of the brane intersection underlying  the AdS$_2$ solutions \eqref{solutionsIIB}. We have presented this solution in Appendix \ref{second}. It is however not obvious to interpret it as associated to a particular brane intersection, for very similar reasons to the ones encountered in \cite{Terrisse:2018hhf}. Still, the interest of the solution is that it describes a black hole geometry that can find interesting applications in the description of four dimensional extremal black holes. For our purposes in this paper, more focused on the defect interpretation of the AdS$_2$ solutions, we will see that the brane intersection depicted in Table \ref{braneintersectionNATD} is fully consistent with the quantised charges associated to the solutions \eqref{NATDyz}, so we will take it as the starting point of our quiver constructions. Note that the solutions described by \eqref{NATDyz} preserve the same supersymmetries of the original class of solutions, since the $S^2$ is left untouched after the non-Abelian T-duality transformation.
\begin{table}[http!]
\renewcommand{\arraystretch}{1}
\begin{center}
\scalebox{1}[1]{
\begin{tabular}{c |c c c  c c  c c c c c}
 branes & $t$ & $x^1$ & $x^2$ & $x^3$ & $z$ & $y$ & $\rho$ & $x^7$ & $x^8$ & $x^9$ \\
\hline \hline
$\mrm{D}4$ & $\times$ & $\times$ & $\times$ & $\times$ & $-$ & $-$ & $\times$ & $-$ & $-$ & $-$ \\
$\mrm{D}6$ & $\times$ & $\times$ & $\times$ & $\times$ & $-$ & $-$ & $-$ & $\times$ & $\times$ & $\times$ \\
$\mrm{NS}5$ & $\times$ & $\times$ & $\times$ & $\times$ & $\times$ & $\times$ & $-$ & $-$ & $-$ & $-$ \\
$\mrm{D}2$ & $\times$ & $-$ & $-$ & $-$ & $-$ & $\times$ & $\times$ & $-$ & $-$ & $-$ \\
$\mrm{F}1$ & $\times$ & $-$ & $-$ & $-$ & $\times$ & $-$ & $-$ & $-$ & $-$ & $-$ \\
$\mrm{D}4'$ & $\times$ & $-$ & $-$ & $-$ & $-$ & $\times$ & $-$ & $\times$ & $\times$ & $\times$ \\
$\mrm{NS}5'$ & $\times$ & $-$ & $-$ & $-$ & $\times$ & $-$ & $\times$ & $\times$ & $\times$ & $\times$ \\
\end{tabular}
}
\end{center}
\caption{BPS/8 intersection describing D2-F1-D4'-NS5' branes ending on D4-D6-NS5 branes, associated to the class of solutions  \eqref{NATDyz}. As before, $x^1,x^2,x^3$ parametrise the directions realising the SO(3) R-symmetry.} \label{braneintersectionNATD}
\end{table}


In the next subsection we turn to the defect interpretation of the solutions.

\subsection{F1-D2-D4'-NS5 line defects within AdS$_5$}

It was shown in  \cite{Lozano:2021fkk} that taking the semi-localised profile defined by
\cite{Youm:1999ti} 
\begin{equation}
H_{\text{D}3}=1+\frac{4\pi q_{\text{D}3}}{(y^2+z^2+r^2)^2},
\end{equation}
a solution in the class given by \eqref{solutionsIIB} is obtained that asymptotes locally to AdS$_5\times S^5$. Taking this same profile in our new class of solutions given by \eqref{NATDyz} and
making the change of coordinates
\begin{equation} \label{changeofcoordinates}
y=\mu \sin{\alpha}\cos{\phi}, \qquad z=\mu \sin{\alpha}\sin{\phi}, \qquad r=\mu \cos{\alpha}
\end{equation}
we again find a solution that asymptotes locally to an AdS$_5$ geometry, in the $\mu\rightarrow 0$ limit. This solution reads\footnote{We have fixed the constants such that the AdS$_5$ subspace has radius one.}
\begin{equation} \label{AdS5NATD}
\begin{split} 
d s_{10}^2 &= \overbrace{\mu^2 \,(ds_{\text{AdS}_2}^2 + ds_{S^2}^2) +  \frac{d\mu^2}{\mu^2}} ^{\text{locally}\,\,\, \text{AdS}_5\,\,\, \text{geometry}} + 
d\alpha^2 + s^2 d\phi^2 + 4 c^{-2} \left(d\rho^2 +  \frac{\rho^2 c^4}{16\, \rho^2 + c^4}  ds_{\tilde S^2}^2 \right) \,, \\
e^{\Phi} &= 2\pi^{-1}q_{\text{D}3}^{-1}c^{-1}(16\, \rho^2 + c^4)^{-1/2}\,, \\
B_{2} &=  \mu s  (\tilde s \text{vol}_{\text{AdS}_2} + \tilde c \text{vol}_{S^2})   +  \frac{16\,\rho^3}{16\,\rho^2+c^4} \, \text{vol}_{\tilde{S}^2}\,,\\
F_{2} &= -2 \pi q_{\text{D}3}  s c^3 d\alpha  \wedge d\phi \,, \\
F_{4} &=  d[4\pi q_{\text{D}3} \rho \mu s (\tilde c \text{vol}_{\text{AdS}_2} - \tilde s \text{vol}_{S^2}) \wedge d\rho]
+  \frac{16\,\rho^3}{16\,\rho^2+c^4} \, \text{vol}_{\tilde{S}^2} \wedge F_{2} +\\
&+2^{-1}\pi q_{\text{D}3} c^3 [(\tilde c \text{vol}_{\text{AdS}_2} - \tilde s \text{vol}_{S^2}) \wedge d\mu \wedge d\alpha + s (\tilde s \text{vol}_{\text{AdS}_2} + \tilde c \text{vol}_{S^2})\wedge  d(\mu c\,d\phi)]\,,\\
F_{6} &= -\frac{4\pi q_{\text{D}3} \rho^3 c^4}{16\, \rho^2 + c^4} \, d[\mu s (\tilde c \, \text{vol}_{\text{AdS}_2} - \tilde s \, \text{vol}_{S^2})] \wedge d[\log(\rho \mu^2 c^2)\, \text{vol}_{\tilde{S}^2}] +\\
&+ 16 \pi q_{\text{D}3} \rho \mu^3\, \text{vol}_{\text{AdS}_2} \wedge \text{vol}_{S^2} \wedge d\mu \wedge d\rho \,.
\end{split}
\end{equation}
Note that in order to obtain this solution we could have alternatively non-Abelian T-dualised the solution in \eqref{solutionsIIB} that asymptotes to AdS$_5\times S^5$, given by equation (3.15) in 
\cite{Lozano:2021fkk}.



The geometry defined by \eqref{AdS5NATD} asymptotes locally to the Gaiotto-Maldacena geometry constructed in \cite{Sfetsos:2010uq}, by acting with non-Abelian T-duality on AdS$_5\times S^5$. 
The fluxes associated to this solution are,
\begin{eqnarray}
&&F_{2}= -2 \pi q_{\text{D}3}  s c^3 d\alpha \wedge d\phi \nonumber\\
&&F_{4}=-2^5\pi  q_{\text{D}3} s c^{-1} \rho^3 H d\alpha \wedge d\phi \wedge \text{vol}_{\tilde S^2}\nonumber\\
&&H_{3}= d\Bigl(\frac{16\,\rho^3}{16\,\rho^2+c^4}\Bigr) \wedge \text{vol}_{\tilde S^2}. \label{fluxesmothertheory}
\end{eqnarray}
The isometries of the AdS$_5$ solution are however broken by the presence of the extra, subleading in $\mu$, fluxes in \eqref{AdS5NATD}, that give rise to new global charges associated to the defect branes. The defect branes backreact then into a geometry described by a 5d curved domain wall with AdS$_2\times S^2$ slicings that is only locally AdS$_5$. The presence of the extra fluxes forbids as well for any supersymmetry enhancement to the 4d $\mathcal{N}=2$ supersymmetry of the AdS$_5$ solution. Note that as in the examples discussed in \cite{Faedo:2020nol,Faedo:2020lyw,Lozano:2022ouq} the R-symmetry of the 4d $\mathcal{N}=2$ AdS$_5$ solution is realised on the internal space, in this case on the ${\tilde S}^2\times S^1_\phi$ subspace, while the R-symmetry of the AdS$_2$ solution becomes part of the superconformal group of the higher dimensional theory.

Let us now proceed with a detailed analysis of the solution described by \eqref{AdS5NATD}.

\section{Line defects within 4d $\mathcal{N}=2$ SCFTs and brane boxes} \label{AdS5part}

In this section we construct the brane set-up associated to the solution \eqref{AdS5NATD} and show that it consists on D2 colour branes stretched between perpendicular NS5-branes. This realises a one dimensional brane box model from which the 1d quiver CFT can be read. We show that this quiver can be embedded within the 4d $\mathcal{N}=2$ SCFT dual to the Gaiotto-Maldacena geometry arising in the asymptotics, described by a linear quiver with gauge groups of increasing ranks \cite{Lozano:2016kum}. Furthermore, we discuss in detail the interpretation of the F1-strings present in the brane intersection, and show that together with the D2 and one of the families of D4-branes describe baryon vertices in the 4d SCFT.

\subsection{The 4d superconformal background theory}

\begin{figure}[h]
\centering
\includegraphics[scale=0.55]{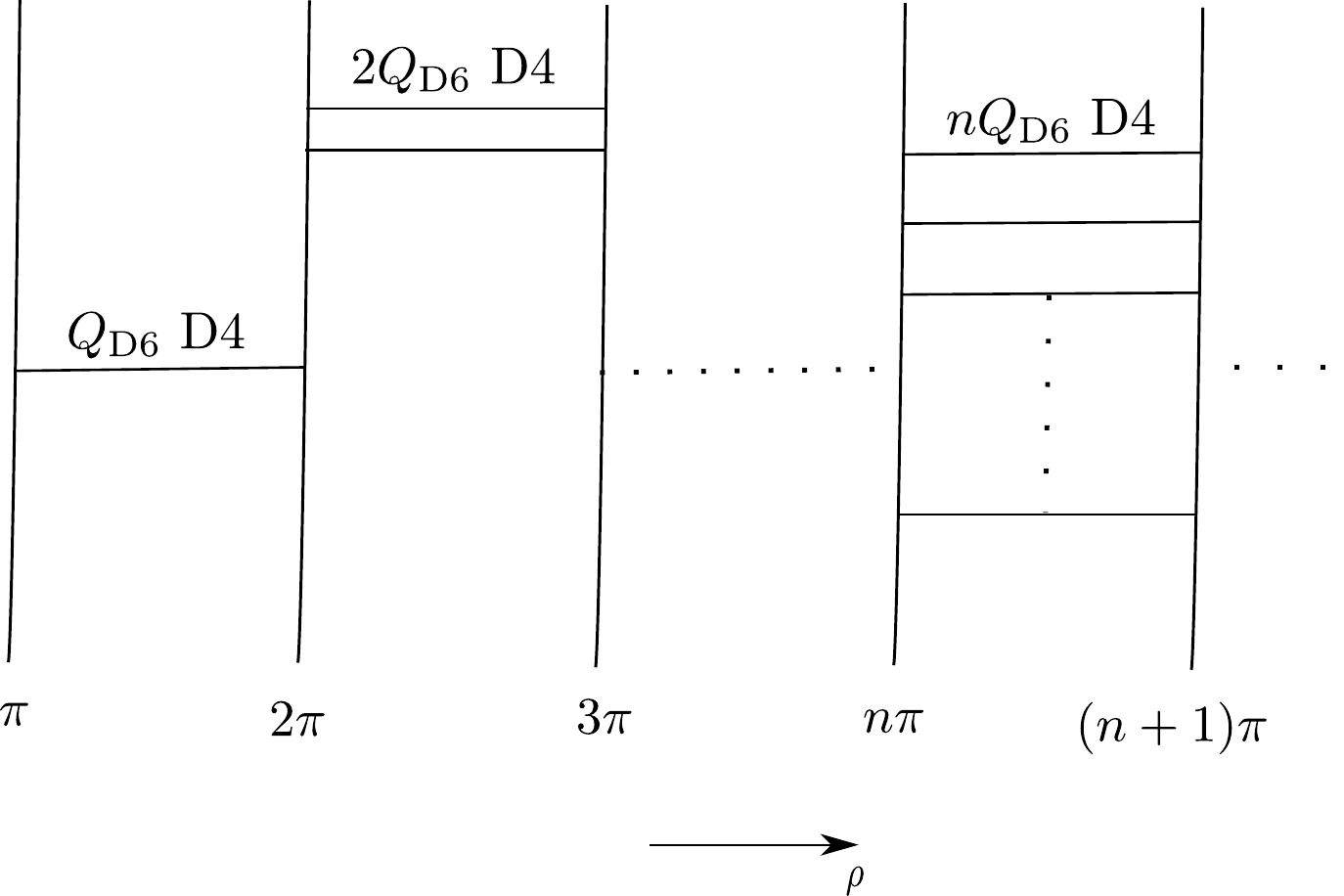}
\caption{Brane set-up associated to the 4d background theory. NS5-branes are located at $\rho_n=n\pi$ and $n Q_{\text{D}6}$ D4-branes are stretched between them in each $[\rho_n,\rho_{n+1}]$ interval.}
\label{4dbranesetup}
\end{figure}

\begin{figure}[h]
\centering
\includegraphics[scale=0.5]{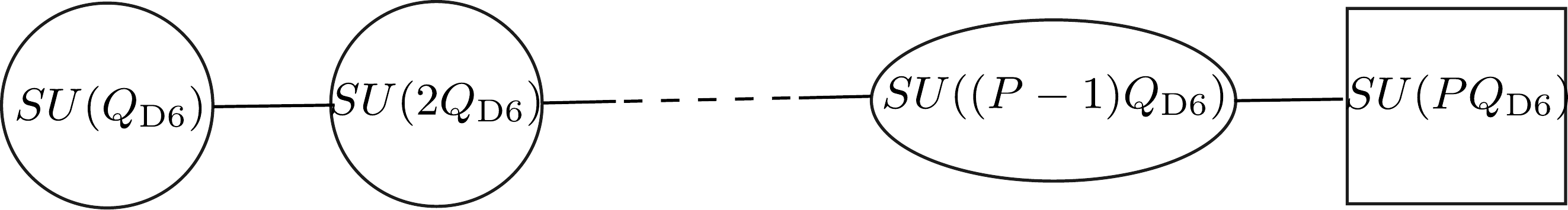}
\caption{Quiver describing the 4d $\mathcal{N}=2$ SCFT dual to the background geometry.} \label{4dquiver}
\end{figure}

As a Gaiotto-Maldacena geometry, the AdS$_5$ solution constructed in  \cite{Sfetsos:2010uq} is dual to a 4d $\mathcal{N}=2$ superconformal field theory living in a D4-NS5-D6 brane intersection. This CFT was thoroughly studied in \cite{Lozano:2016kum}, to which the reader is referred for more details. The D4-NS5-D6 intersection is the one depicted in the first three lines in Table \ref{braneintersectionNATD}.
We start computing the quantised charges associated to the NS5-branes, following \cite{Lozano:2016kum}. Integrating 
\begin{equation}
B_2^{{\tilde S}^2}=\frac{16\,\rho^3}{16\,\rho^2+c^4}\text{vol}_{{\tilde S}^2}
\end{equation}
on the cycle defined by the ${\tilde S}^2$ positioned at $\alpha=\pi/2$, we have,
\begin{equation}\label{cond}
\frac{1}{(2\pi)^2}\int_{\Sigma_2}B_2^{{\tilde S}^2}=\frac{\rho}{\pi}.
\end{equation}
Since this quantity has to take values between 0 and 1 in order to have a well-defined partition function, the $\rho$ direction must be divided in intervals of length $\pi$, such that when $\rho\in [n\pi,(n+1)\pi]$ a large gauge transformation of parameter $n$ must be performed for \eqref{cond} to be satisfied. $B^{{\tilde S}^2}_{2}$ must thus be modified as
\begin{equation}\label{B2mo}
B^{{\tilde S}^2}_{2}=\Bigl(\frac{16\,\rho^3}{16\,\rho^2+c^4}-n\pi\Bigr)\text{vol}_{{\tilde S}^2},
\end{equation}
for $\rho\in [n\pi,(n+1)\pi]$. One effect of this split into intervals is that upon crossing $\rho=n\pi$ a NS5-brane is created, generating a Hanany-Witten brane creation effect. Indeed, we have in each interval
\begin{equation}
Q_{\text{NS}5}=\frac{1}{(2\pi)^2}\int_{\Sigma_3} H_3=\frac{1}{(2\pi)^2}\int_{n\pi}^{(n+1)\pi}\int_{\Sigma_2}H_3=1,
\end{equation}
with $\Sigma_2$ the ${\tilde S}^2$ located at $\alpha=\pi/2$. Moreover, the $n$ term in \eqref{B2mo} contributes to the 4-form Page flux with ${\tilde S}^2$ component such that
\begin{equation}
\hat{F}_{4}^{{\tilde S}^2}=n\pi\, \text{vol}_{{\tilde S}^2}\wedge F_{2},
\end{equation}
which using\footnote{We use units in which $g_s=\alpha'=1$.}
\begin{equation}
Q_{\text{D}p}=\frac{1}{(2\pi)^{7-p}}\int_{\Sigma_{8-p}}\hat{F}_{8-p},
\end{equation}
gives, in the $\rho\in [n\pi, (n+1)\pi]$ interval,
\begin{equation}
Q_{\text{D}4}=n\, Q_{\text{D}6},
\end{equation}
where\footnote{Note that as usual the quantised charges need to be renormalised after a non-Abelian T-duality transformation. This can be done through a redefinition of Newton's constant. This will affect our normalisation of the holographic central charge in section \ref{centralcharge} (see for instance \cite{Lozano:2016kum}).}
\begin{equation}\label{QD6}
Q_{\text{D}6}=\frac{\pi}{2}q_{\text{D}3}.
\end{equation}
These quantised charges give rise to the Hanany-Witten brane set-up depicted in Figure \ref{4dbranesetup}, in which the D4-branes play the role of colour branes, and there are no D6 flavour branes as the D6-brane charge remains constant across intervals.
Note that this brane set-up extends infinitely in the $\rho$-direction, due to the non-compact character of the $\rho$-coordinate. This happens because after the non-Abelian T-duality transformation the $S^3$ of the original solution is replaced by an $\mathbb{R}^3$ subspace, and due to our lack of knowledge of how non-Abelian T-duality extends beyond spherical worldsheets it is not possible to infer its global properties \cite{Alvarez:1993qi}.
 In view of this in \cite{Lozano:2016kum} different ways of terminating the brane set-up were discussed. Here we will  choose the simplest scenario, namely, we will terminate the brane set-up at $\rho_P=P\pi$ by adding a set of $PQ_{\text{D}6}$ D6-branes (or semi-infinite D4-branes). The resulting quiver is the one depicted in Figure  
\ref{4dquiver}. 
One can check that at each node of the quiver the condition on the ranks of the gauge groups, $l_i$,
\begin{equation}
2l_n=l_{n+1}+l_{n-1},
\end{equation}
required for the $\beta$-function to vanish  \cite{Witten:1997sc}, is satisfied. Moreover the field theory and holographic central charges can be shown to agree to leading order in $P$ (the holographic limit in these quiver constructions)  \cite{Lozano:2016kum}.

\subsection{The line defect theory}\label{braneboxes}

We proceed now to the description of the D2-F1-D4'-NS5' defect branes whose backreaction in the AdS$_5$ geometry generates the AdS$_2$ solution. We start focusing on the F1- and NS5' branes. For this purpose it is useful to define
\begin{equation}\label{mu}
y=\mu \sin{\alpha}\cos{\phi}, \qquad z=\mu \sin{\alpha}\sin{\phi},
\end{equation}
as in equation \eqref{changeofcoordinates}.
The $B_2$ field then reads
\begin{equation}\label{B2field}
B_2=z\,\text{vol}_{\text{AdS}_2}+y\,\text{vol}_{S^2}+\frac{16\,\rho^3}{16\,\rho^2+c^4} \text{vol}_{{\tilde S}^2}.
\end{equation}
The component along the ${\tilde S}^2$ is associated, as we have shown, to the NS5-branes of the 4d background theory, so we will no longer discuss it. In turn, the first and second components are associated to the F1 and NS5' defect branes. Let us start looking at the NS5'-branes. A very similar analysis to the one performed previously for the NS5-branes of the background theory shows that we must divide the $y$-direction in $[m\pi,(m+1)\pi]$ intervals and perform a large gauge transformation of gauge parameter $m$ in each one of them to satisfy that $B_2$ lies in the fundamental region. This 
fixes
\begin{equation}
B_2^{S^2}=(y-m\pi)\text{vol}_{S^2},
\end{equation}
in these intervals, and creates NS5'-branes along the $y$ direction at $y=m\pi$. Let us now turn our attention to the electric component of $B_2$.
The F1-strings are electrically charged with respect to the NS-NS 3-form. Therefore, their charges are computed according to
\begin{equation}
Q_{\text{F}1}^e=\frac{1}{(2\pi)^2}\int H_3=\frac{1}{(2\pi)^2}\int_{\text{AdS}_2} B_2^e,
\end{equation}
where we use the superscript $e$ to denote that we are referring to electric as opposed to magnetic charges.
Regularising the volume of the AdS$_2$ space such that $\text{Vol}_{\text{AdS}_2}=4\pi$ \footnote{This regularisation prescription is based on the analytical continuation that relates the AdS$_2$ space with an $S^2$.} and dividing the $z$ direction into intervals of length $[k\pi,(k+1)\pi]$ a F1-string lies in each such interval. This is also implied by the condition that the integral of $B_2$ lies in the fundamental region, as previously discussed (see \cite{Lozano:2020sae}) . In this case a large gauge transformation of gauge parameter $k$ must be performed for $z\in [k\pi,(k+1)\pi]$, such that
\begin{equation}\label{B2electric}
B_2^e=(z-k\pi)\text{vol}_{\text{AdS}_2}
\end{equation}
in this interval. We will come back to the physical interpretation of this condition after we discuss the quiver quantum mechanics associated to the solution.

\begin{figure}[H]
\centering
\includegraphics[scale=0.6]{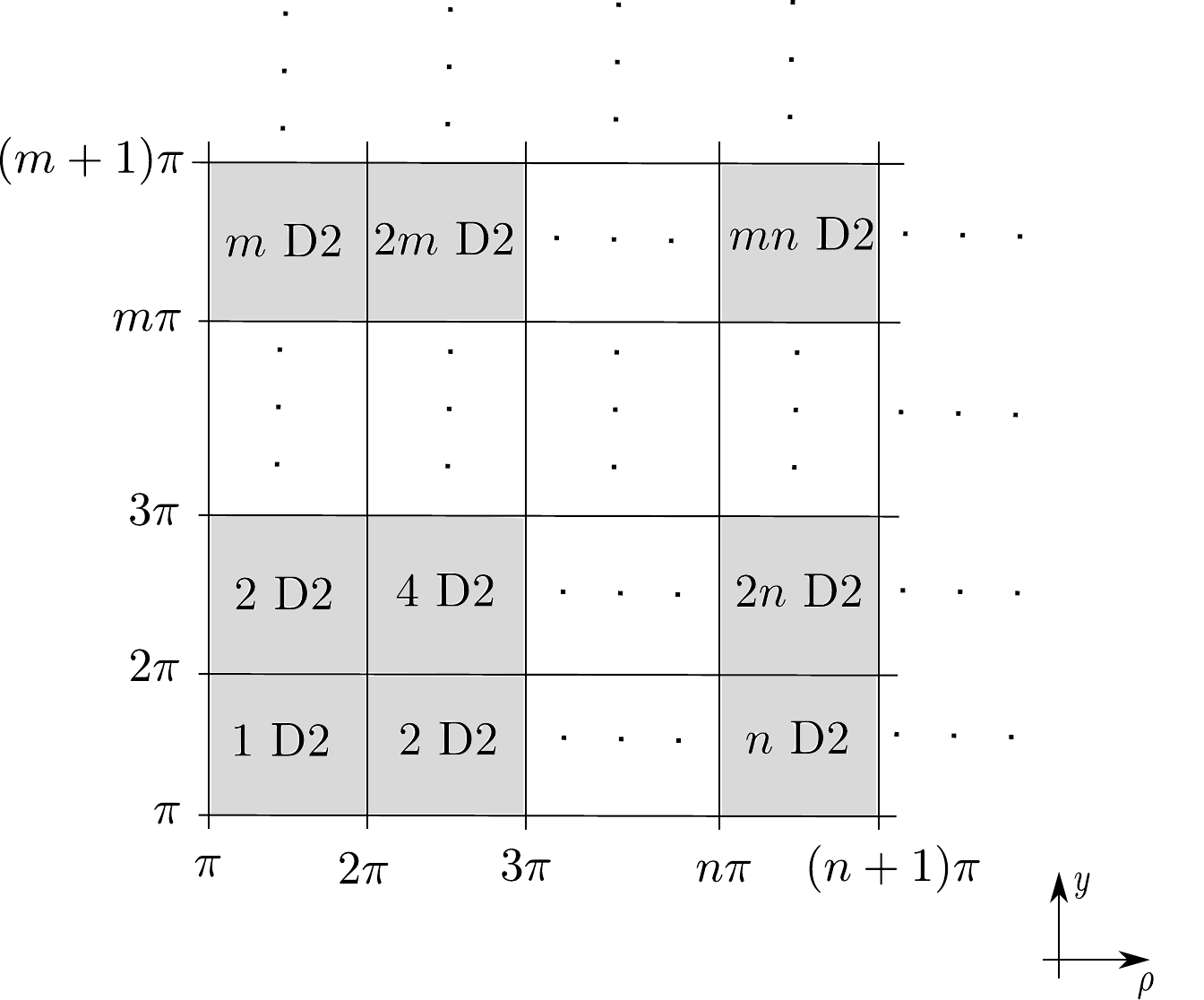}
\caption{Brane set-up associated to the AdS$_2$ solution  \eqref{AdS5NATD} (in units of $Q_{\text{D}6}=1$).} \label{branesetup}
\end{figure}

For this purpose let us now look at the D2 and D4' defect branes. The large gauge transformations of parameters $m$ and $k$ modify the $S^2$ and $\text{AdS}_2$ components of the Page fluxes, according to
\begin{eqnarray}
&&{\hat F}_4\rightarrow {\hat F}_4+m\pi F_2\wedge \text{vol}_{S^2} +k\pi F_2\wedge \text{vol}_{{\text{AdS}_2}}\,,\\
&&{\hat F}_6\rightarrow {\hat F}_6+mn\pi^2 F_2\wedge \text{vol}_{S^2}\wedge \text{vol}_{{\tilde S}^2}+kn\pi^2 F_2\wedge \text{vol}_{\text{AdS}_2}\wedge \text{vol}_{{\tilde S}^2}\,,
\end{eqnarray}
generating a Hanany-Witten brane creation effect. This affects the $(\alpha,\phi)$ components of the Page fluxes, where $F_2$ lies.
Since in the absence of large gauge transformations we have for these components that
\begin{equation}
F_4=F_2\wedge B_2\,, \qquad F_6=F_4\wedge B_2-\frac12 F_2\wedge B_2\wedge B_2\,,
\end{equation}
we find that
\begin{eqnarray}
&&{\hat F}_4 = k\pi F_2\wedge \text{vol}_{{\text{AdS}_2}} + m\pi F_2\wedge \text{vol}_{S^2} + n\pi F_2\wedge \text{vol}_{\tilde S^2}\,, \label{creationD4'}\\
&&{\hat F}_6= mn\pi^2 F_2\wedge \text{vol}_{S^2}\wedge \text{vol}_{{\tilde S}^2}+kn\pi^2 F_2\wedge \text{vol}_{\text{AdS}_2}\wedge \text{vol}_{{\tilde S}^2}\,.\label{creationD2}
\end{eqnarray}
Let us focus on the magnetic components. Equations \eqref{creationD4'}, \eqref{creationD2} imply
that D4' branes are created across NS5'-branes as we move in the $y$-direction, and D2-branes are created both across NS5'-branes as we move in the $y$-direction and across NS5-branes as we move in the $\rho$-direction. To this we have to add the D4-branes that were already created across NS5-branes in the $\rho$-direction in the background theory. 
The corresponding quantised charges in the $y\in [m\pi,(m+1)\pi]$, $\rho\in [n\pi,(n+1)\pi]$ intervals are given by
\begin{equation} \label{magneticcharges}
Q_{\text{D}4'}=m\, Q_{\text{D}6}, \qquad Q_{\text{D}2}=mn\, Q_{\text{D}6}, \qquad Q_{\text{D}4}=n\, Q_{\text{D}6}.
\end{equation}
We thus find a brane scenario in which two directions, $y$ and $\rho$, play the role of field theory directions. Being the D2-branes stretched between both types of NS5 and NS5' branes in these directions they are interpreted as the colour branes where a 1d supersymmetric field theory lives.
In turn, the charges carried by the D4' and D4 branes are induced by the D2-branes they end on, with which they share, respectively, the $y$ and $\rho$ field theory directions (see Table \ref{braneintersectionNATD}). The brane set-up in the $(\rho,y)$ plane is then the one depicted in Figure \ref{branesetup}. 
Once we have identified the brane set-up we can proceed to construct the quiver mechanics that describes the field theory living in the D2-branes. In order to do that we must look at the quantisation of the open strings that stretch between the branes in the different boxes. As it is customary for 1d $\mathcal{N}=4$ multiplets we will use 2d $(0,4)$ notation. We will follow closely \cite{Hanany:2018hlz}, where the quantisation of open strings in D3-brane box models realising 2d (0,4) field theories was studied in detail. Our brane set-up is simply related to the D3-NS5-D5-NS5'-D5' brane intersection studied in \cite{Hanany:2018hlz} by (Abelian) T-duality along the $x^1$ direction therein, and thus realises a 1d $\mathcal{N}=4$ instead of a 2d (0,4) field theory.  Other than this the analysis is completely analogous. There are four types of D2-D2 strings to consider:

\begin{itemize}
\item When the end-points of the string lie on the same stack of D2-branes the projections induced by both the NS5 and the NS5' branes leave behind a (0,4) vector multiplet, since the D2-branes cannot move in any of the transverse directions. 

\item When the end-points of the string lie on two different stacks of D2-branes separated by an NS5-brane the degrees of freedom along the $(x^7,x^8,x^9)$ directions are fixed, leaving behind the scalars associated to the $(x^1,x^2,x^3)$ directions, which together with the $A_y$ component of the gauge field give rise to a (0,4) twisted hypermultiplet in the bifundamental representation, since the scalars are charged under the R-symmetry. 

\item When the end-points of the  string lie on two different stacks of D2-branes separated by an NS5'-brane the degrees of freedom along the $(x^1,x^2,x^3)$ directions are fixed, leaving behind the scalars associated to the $(x^7,x^8,x^9)$ directions, which together with the $A_{\rho}$ component of the gauge field give rise to a (0,4) hypermultiplet in the bifundamental representation, since the scalars are uncharged under the R-symmetry. 

\item When the end-points of the string lie on two different stacks of D2-branes separated by both an NS5 and an NS5' brane all the scalars are fixed, leaving behind the fermionic mode associated to a bifundamental (0,2) Fermi multiplet. 



\end{itemize}

\begin{figure}[H]
\centering
\includegraphics[scale=0.6]{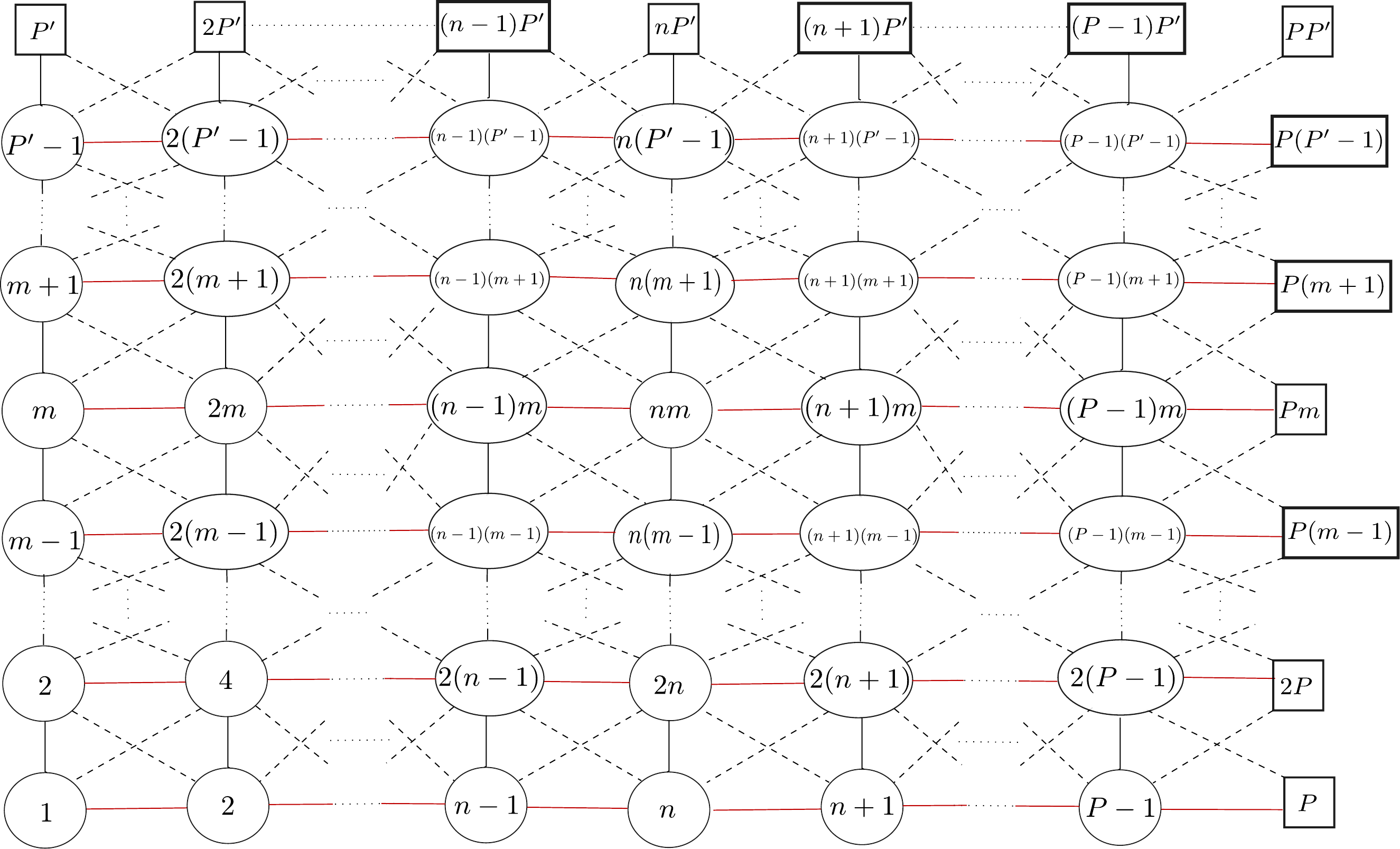}
\caption{Quiver quantum mechanics associated to the AdS$_2$ solution  given by \eqref{AdS5NATD}. Circles denote (0,4) vector multiplets, red lines (0,4) bifundamental twisted hypermultiplets, black lines (0,4) bifundamental hypermultiplets and dashed lines bifundamental (0,2) Fermi multiplets. We have taken units in which $Q_{\text{D}6}=1$.}\label{branebox}
\end{figure}

The previous fields give rise to the planar quiver depicted in Figure \ref{branebox}. This quiver consists of two arrows of linear quivers, associated to the D2-branes stretched between NS5-branes in the $\rho$ direction and NS5'-branes in the $y$ direction, with mutual interactions consisting of (0,2) Fermi multiplets. The quiver is terminated in both directions with two families of flavour groups, arising from $n P'$ D6-branes (or semi-infinite D4'-branes) placed at $\rho=n\pi$, with $n=1,2,\dots, P$, $y_{P'}=P'\pi$ and $m P$ D6-branes (or semi-infinite D4-branes) placed at $\rho_P=P\pi$, $y=m\pi$ with $m=1,2,\dots, P'$. This allows to construct a well-defined one dimensional quiver quantum mechanics from which we can compute the degrees of freedom of the 1d SCQM that arises in the IR, as we will pursue in section \ref{centralcharge}.
Our proposal is that the quiver depicted in Figure \ref{branebox} describes a 1d field theory that flows in the IR to the 1d SCQM dual to the AdS$_2$ solution. Note that as a one dimensional field theory there is no condition for gauge anomaly cancelation. Yet, it is striking that the quiver mechanics satisfies the conditions for gauge anomaly cancelation of a 2d (0,4) field theory. This can be checked straightforwardly in our 2d (0,4) notation for the superfields. For this we need to recall the contribution from the different fields to the U(N) gauge anomaly (see for instance  \cite{Putrov:2015jpa,Hanany:2018hlz,Lozano:2019zvg,Couzens:2021veb}). This contribution is as follows. (0,4) twisted or untwisted hypermultiplets contribute with a factor 2N if they are in the adjoint or with a factor 1 if they are in the fundamental representation. (0,4) vector multiplets contribute with a factor -2N. (0,2) Fermi multiplets in the fundamental contribute with a factor -1/2. With these contributions one can easily check that all nodes in the quiver in Figure \ref{branebox} satisfy that the gauge anomaly vanishes. We will discuss further this property of the quiver when we interpret  our result for the central charge in section \ref{centralcharge}.

\begin{figure}[H]
\centering
\includegraphics[scale=0.6]{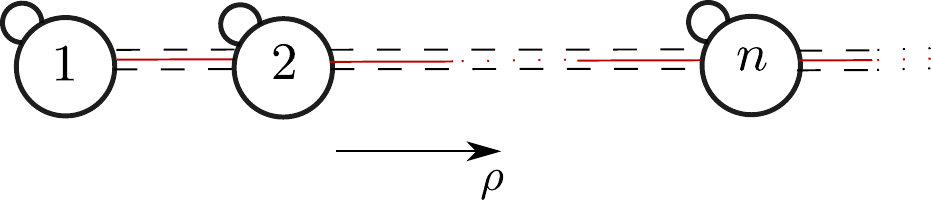}
\caption{Field theory living in the D4-NS5-D6 subsystem (in units of $Q_{\text{D}6}=1$) in terms of 2d (0,4) multiplets.} \label{D2-chain-rho}
\end{figure}

Another interesting property of the quiver is that it can be seen as the result of embedding the D2-D4'-NS5'-F1 defect branes in the 4d $\mathcal{N}=2$ SCFT living in the D4-NS5-D6 brane intersection. Let us describe how this happens.  From the point of view of the 4d theory the R-symmetry is realised on the $x^7,x^8,x^9$ directions, as mentioned in section \ref{Newclass}. The 4d quiver depicted in Figure \ref{4dquiver} is then decomposed in terms of 2d (0,4) matter fields as shown in Figure \ref{D2-chain-rho}.  
\noindent In this decomposition the 4d
$\mathcal{N}=2$ vector multiplet gives rise to a (0,4) vector multiplet with gauge field $A_\alpha$, $\alpha=t,z$, plus a (0,4) adjoint hypermultiplet, arising from
combining the reduction of the gauge field along the $(x^1,x^2,x^3)$ directions and the fluctuations in the $y$-direction. In turn, the 4d $\mathcal{N}=2$ bifundamental hypermultiplet gives rise to a (4,4) bifundamental twisted hypermultiplet, arising from combining the $A_\rho$ component of the gauge field with the fluctuations in $(x^7,x^8,x^9)$. 

\begin{figure}[H]
\centering
\includegraphics[scale=0.6]{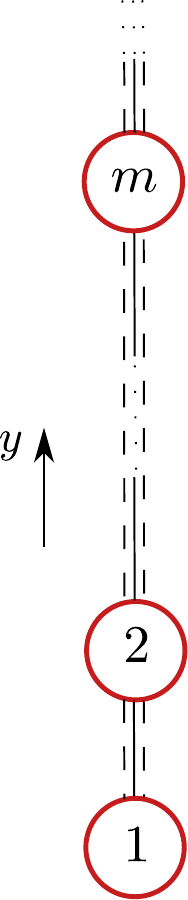}
\caption{Field theory living in the D4'-NS5'-D6 subsystem (in units of $Q_{\text{D}6}=1$) in terms of 2d (0,4) multiplets. Red circles represent (4,4) vector multiplets.} \label{D2-chain-y}
\end{figure}

We can consider in an analogous way the D4'-NS5'-D6 brane subsystem of the brane set-up shown in Table \ref{braneintersectionNATD}. The 4d SCFT living in this brane subsystem is again the one depicted in Figure \ref{4dquiver}, with $y$ playing now the role of field theory direction. However, in the decomposition into 2d  (0,4) matter multiplets the 4d $\mathcal{N}=2$ vector multiplet decomposes into a (0,4) vector multiplet plus a (0,4) adjoint twisted hypermultiplet, arising from combining the reduction of the gauge field along the $(x^7,x^8,x^9)$ directions and the fluctuations in the $\rho$-direction. As it is well-known these multiplets combine into a (4,4) vector multiplet. The difference with the decomposition of the gauge field living in the D4-branes is that the scalars are now charged with respect to the SU(2) R-symmetry. In turn, the 4d $\mathcal{N}=2$ bifundamental hypermultiplet gives rise to a (4,4) bifundamental hypermultiplet, arising from combining the $A_y$ component of the gauge field with the fluctuations in $(x^1,x^2,x^3)$. Again, the difference with the decomposition of the 4d hypermultiplet living on the D4-branes is that the scalars are now uncharged with respect to the SU(2) R-symmetry. The 4d quiver living in the D4'-NS5-D6 branes is represented in terms of 2d (0,4) multiplets in Figure \ref{D2-chain-y}. 
The quiver depicted in Figure \ref{branebox} can now be seen as the result of assembling the two quivers represented in Figures \ref{D2-chain-rho} and \ref{D2-chain-y}. In this assembly the charges carried by the D4 and D4'-branes are now carried by D2-branes, that stretch in both the $\rho$ and $y$ field theory directions. Our proposal is that the 1d field theory described by this quiver flows in the IR to the SCQM dual to the AdS$_2$ solution \eqref{AdS5NATD}. We would like to stress that the SCQM proposed in this section is far more elaborated than those previously constructed in the literature \cite{Assel:2018rcw,Assel:2019iae,Lozano:2020txg,Lozano:2020sae,Lozano:2021rmk,Ramirez:2021tkd,Lozano:2021fkk}, since it involves the highly non-trivial brane box models constructed in \cite{Hanany:2018hlz}, now realising an $\mathcal{N}=4$ supersymmetric quantum mechanics. Moreover, we have at our disposal the explicit holographic dual, and therefore a well-controlled string theory realisation that allows to study these constructions geometrically. We will see that this is particularly useful when addressing the, non-trivial issue, of the computation of the central charge. Our construction provides, to our knowledge, the first example in which a brane box model has been described holographically\footnote{See \cite{Faedo:2020lyw} for AdS$_3$ solutions dual to D3- brane boxes with one circular direction.}. We would like to emphasise the non-trivial role played by non-Abelian T-duality in making this possible.

In the next section we turn to a more precise interpretation of the massive F1-strings present in the defect sector of the theory. Our discussion  follows very closely the analysis carried out in section 4.6 in \cite{Lozano:2021fkk}. In that reference a  brane set-up like the one shown in Table \ref{braneintersectionNATD} was studied in the particular case in which the $\rho$ and $y$ directions are circular and the D4 and the D6-branes are smeared along them. This gave rise to a class of AdS$_2$ solutions that were missing an AdS$_5$ asymptotics. Yet the analysis of the field theory dual allowed to interpret the D2-D4'-F1 branes as baryon vertices for the D4-D6-NS5 subsystem, suggesting that a defect interpretation should still be possible. In the next section we show that the previous baryon vertex interpretation goes through, as expected, for our AdS$_2$/CFT$_1$ set-up, which finds in this way a defect interpretation from both the field theory and the geometrical sides.

\subsection{Baryon vertex interpretation}

In this section we turn to the interpretation of the F1-strings of the solution. We show that together with the D2 and the D4' branes they find a baryon vertex interpretation within the 4d $\mathcal{N}=2$ background theory.

\begin{figure}[H]
\centering
\includegraphics[scale=0.55]{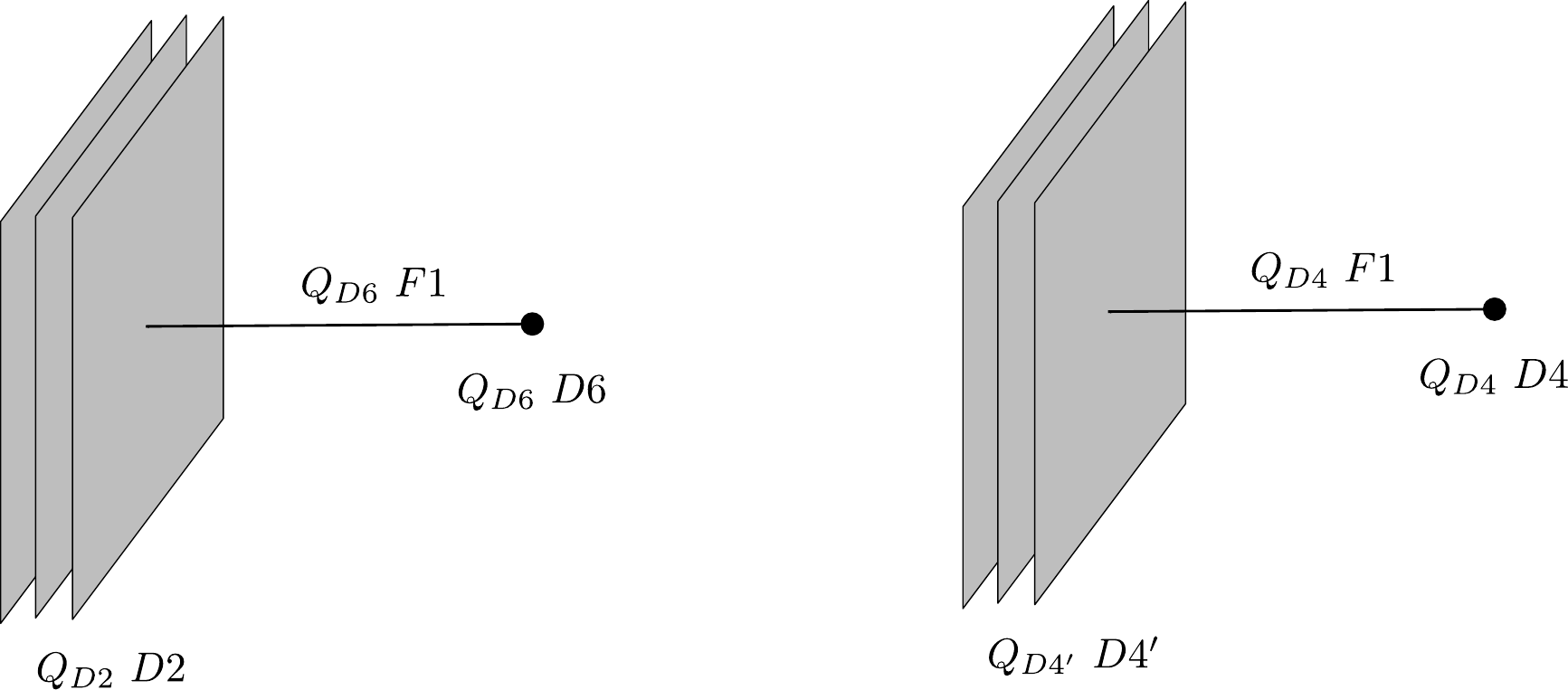}
\caption{Wilson loop in the $Q_{\text{D}6}$ ($Q_{\text{D}4}$) antisymmetric representation of U($Q_{\text{D}2}$) (U($Q_{\text{D}4'}$)).} \label{buildingblock}
\end{figure}

\begin{figure}[H]
\centering
\includegraphics[scale=0.6]{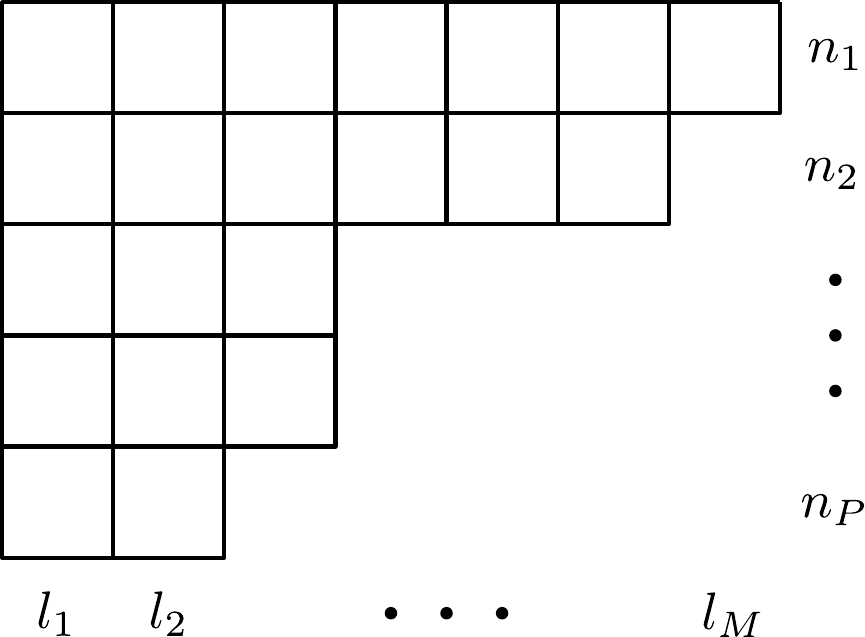}
\caption{Young tableau labelling the irreducible representations of U(N).}\label{youngtableau}
\end{figure}

\begin{figure}[H]
\centering
\includegraphics[scale=0.6]{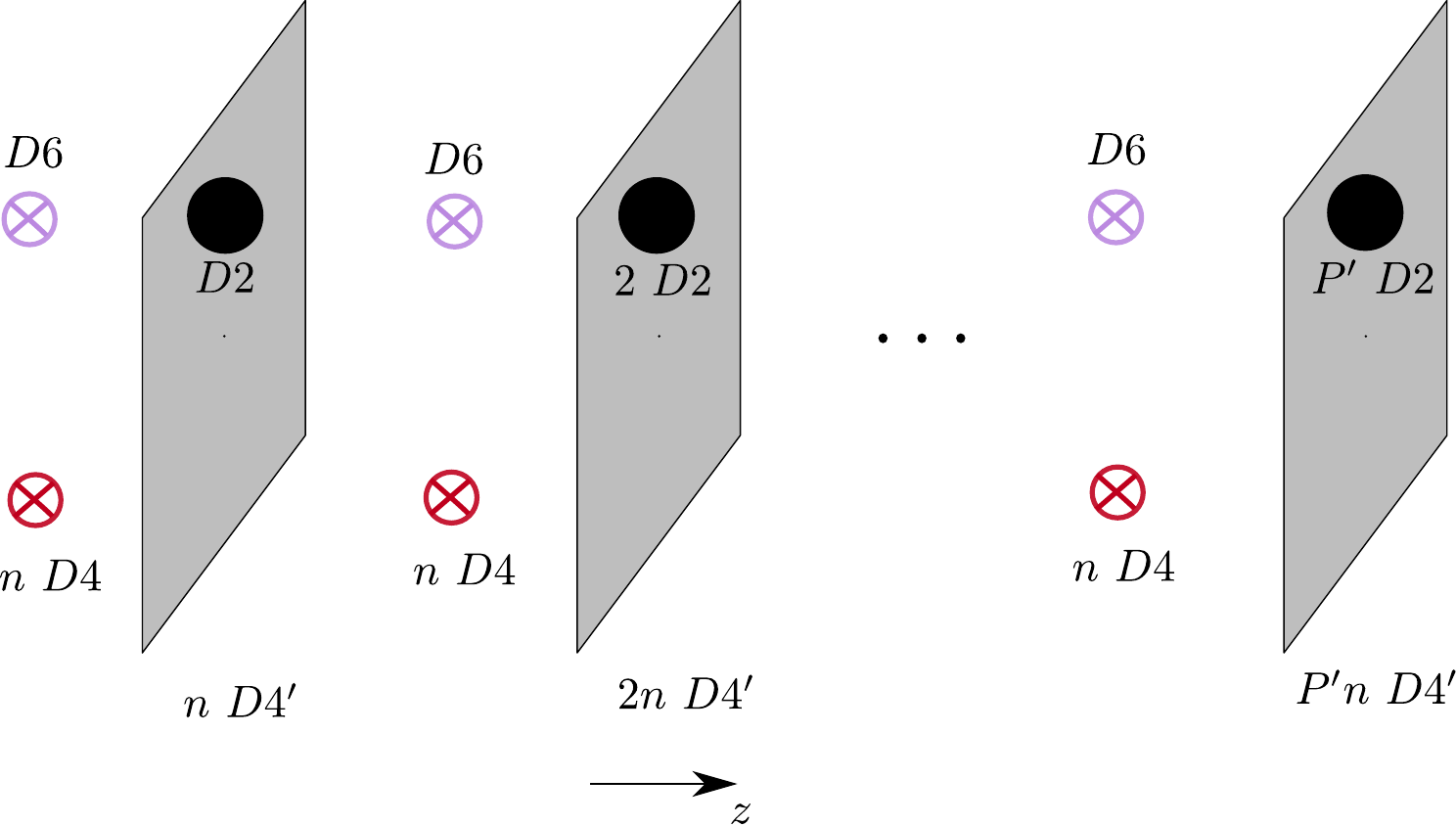}
\caption{Brane set-up in the $z$-direction, for $y$ and $\rho$ constants. The numbers of D4 and D6 branes at each interval are given by their respective magnetic charges. Instead, for the numbers of D2 and D4' branes we give their electric charges (computed in \eqref{D2electric} and \eqref{D4'electric}) as these are the ones that play a role in their interpretation as baryon vertices.}\label{D2-D4-D4'-D6}
\end{figure}

\begin{figure}[H]
\centering
\includegraphics[scale=0.6]{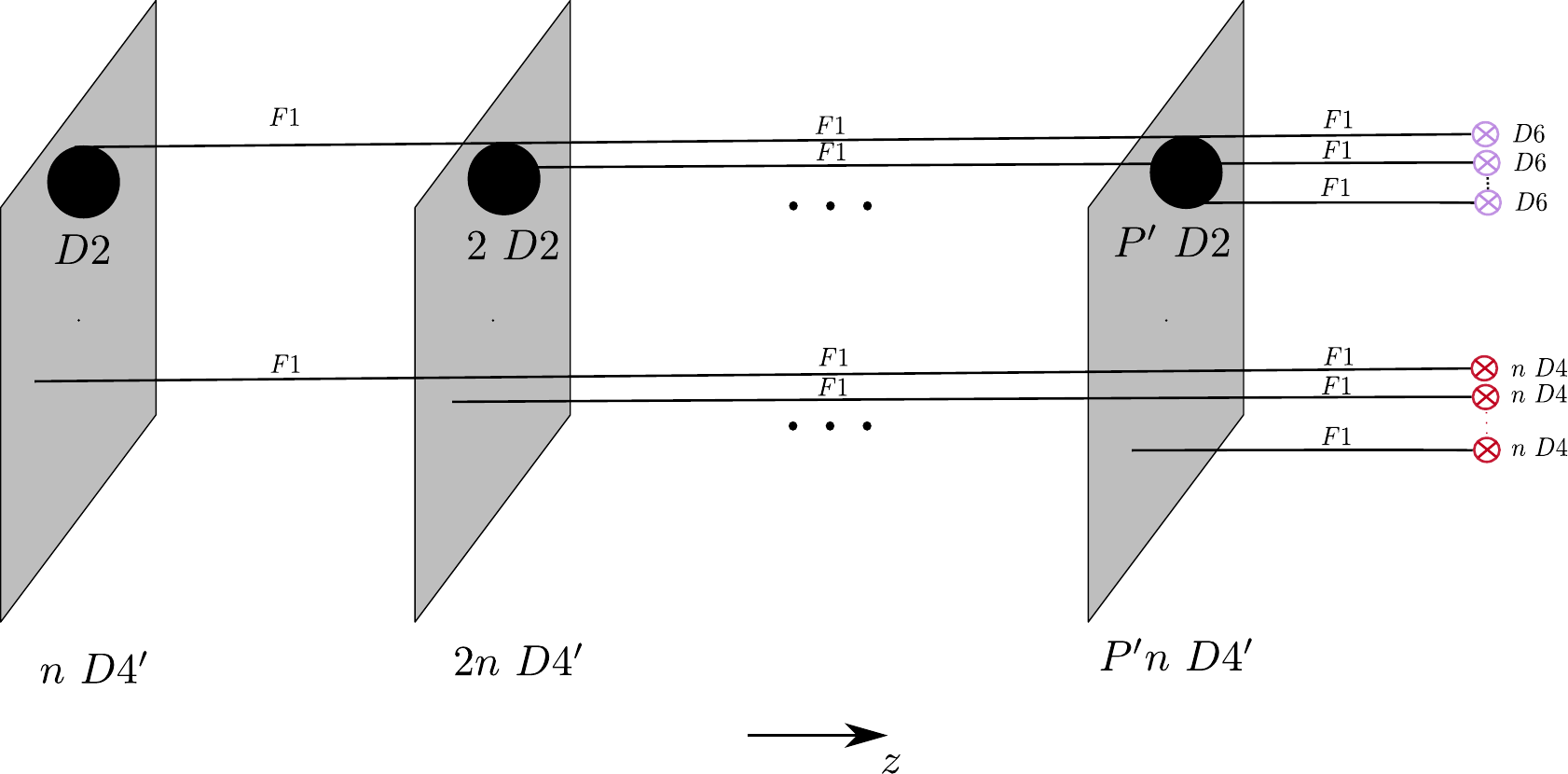}
\caption{Hanany-Witten brane set-up equivalent to the brane configuration in Figure \ref{D2-D4-D4'-D6}.}\label{D2-D4-D4'-D6-F1}
\end{figure}

Let us start  looking at the D2-branes. The following worldvolume coupling
\begin{equation}
S_{\text{D}2}=T_2\int F_2\wedge A_t,
\end{equation}
shows that a D2-brane lying on $(t,\phi, \alpha)$ behaves as a baryon vertex for the D6-branes, since it carries $Q_{\text{D}6}$ units of F1-string charge. Analogously, the coupling
\begin{equation}
S_{\text{D}4'}=T_4\int \hat{F}_4\wedge A_t
\end{equation}
in the worldvolume of a D4'-brane shows that a D4'-brane lying on $(t,\phi,\alpha,{\tilde S}^2)$ and located at a fixed position in the $\rho\in [n\pi,(n+1)\pi]$ interval behaves as a baryon vertex for the D4-branes lying in this interval, since it carries $Q_{\text{D}4}=nQ_{\text{D}6}$ units of F1-string charge. 
Indeed, the relative orientation between the D4' and the D4 branes, and between the D2 and the D6 branes in the brane set-up is the one that allows to create F1-strings stretched between the D4' and the D4 and the D2 and the D6 branes, as depicted in Figure \ref{buildingblock}. 
\noindent These strings have as their lowest energy excitation a fermionic field, which upon integration leads to a Wilson loop. It was shown in \cite{Yamaguchi:2006tq,Gomis:2006sb} that in order to describe a half-BPS Wilson loop in an antisymmetric representation labelled by a Young tableau (depicted in Figure \ref{youngtableau}) one must consider a configuration of stacks of branes separated a distance $L$ from the colour branes, with $(l_1,l_2,\dots, l_M)$ F1-strings stretched between the stacks.
This generalises the description of a Wilson loop in \cite{Rey:1998ik,Maldacena:1998im} to all other antisymmetric representations\footnote{See \cite{Gomis:2006sb,Gomis:2006im} for a similar description of half-BPS Wilson loops in symmetric representations.}.
Coming back to our brane configuration the relevant fluxes that give the quantised electric charges of the D2 and D4' branes playing the role of baryon vertices are the $(\text{AdS}_2,\phi,\alpha)$ and $(\text{AdS}_2,\phi,\alpha,{\tilde S}^2)$ components of $\hat{F}_4$ and $\hat{F}_6$ found in \eqref{creationD4'} and \eqref{creationD2}, given by
\begin{align}
\begin{split}
\hat{F}_4^e=& \,k\pi \, F_2 \wedge \text{vol}_{\text{AdS}_2} \\
\hat{F}_6^e=& \, n k \pi^2 \, F_2 \wedge \text{vol}_{\text{AdS}_2}\wedge \text{vol}_{{\tilde S}^2}.
\end{split}
\end{align}
These fluxes give rise to the quantised charges
\begin{equation}\label{D2electric}
Q_{\text{D}2}^e=k\, Q_{\text{D}6},
\end{equation}
for $z\in [k\pi,(k+1)\pi]$, and
\begin{equation}\label{D4'electric}
Q_{\text{D}4'}^e=n\, Q_{\text{D}2}^e=nk\, Q_{\text{D}6},
\end{equation}
for $z\in [k\pi,(k+1)\pi]$ and $\rho\in [n\pi,(n+1)\pi]$. 
From these charges we can read the brane set-up along the $z$ direction for constant $y$ and $\rho$, that we depict in Figure \ref{D2-D4-D4'-D6}. In this set-up the sets of D6-branes located at $\rho=n\pi$, $y_{P'}=P'\pi$ that terminated the quiver quantum mechanics in the $y$ direction allow us to also terminate the brane set-up in the $z$ direction, if we locate them at 
$z_{P'}=P'\pi$ \footnote{Note that the relations \eqref{mu} imply that $z$ and $y$ must reach the same maximum values.}.
The brane set-up depicted in Figure \ref{D2-D4-D4'-D6} can now be related by a combination of a T-duality, an S-duality, successive Hanany-Witten moves and a further T-duality to the brane set-up depicted in Figure \ref{D2-D4-D4'-D6-F1}. This is carefully explained in  \cite{Lozano:2020sae} (see also \cite{Lozano:2021fkk}).
In this description of the system the relation with the constructions in \cite{Yamaguchi:2006tq,Gomis:2006sb}   becomes manifest.
In our case the sum of the F1-strings stretched between each D2 (D4') and the flavour D6 (D4) branes coincides with the rank of the gauge group of the D2 (D4') branes. This implies that the Wilson lines are in the fundamental representation of the gauge groups. Therefore, the D2-D4' branes describe baryon vertices for the D6-D4 branes of the 4d background theory. As we move in the $\rho$ direction the NS5-branes located in the different positions in $\rho$ allow to create D4-branes stretched between them in an increasing number in units of $Q_{\text{D}6}$. Exactly the same phenomenon takes place for the D4'-branes, which are created, orthogonal to the D4-branes, as the NS5-branes are crossed, in an increasing number in units of $Q_{\text{D}2}$, since they carry electric charge. In turn, as the number of D2-branes varies as we move in the $z$ direction, the same happens with the D4'-branes. In this way one finds an analogous interpretation to that of the D2 branes for the D4'-branes, as baryon vertices for the D4-branes. 

The conclusion of our analysis in this section is that the AdS$_2$ solution can be interpreted as describing backreacted baryon vertices within the 4d $\mathcal{N}=2$ CFT living in the D4-NS5-D6 branes. Consistently with this interpretation the AdS$_5$ solution associated to the D4-NS5-D6 intersection arises asymptotically locally far away from the defect. Even if the baryon vertices are described in terms of D2-D4'-F1 branes, NS5'-branes need also be introduced in the background such that a solution to Type IIA supergravity arises. The full brane set-up allows then for a field theory description of the AdS$_2$ solution in terms of D2-branes stretched between both the NS5-branes and the NS5'-branes in two perpendicular directions. 

\subsection{Computation of the central charge} \label{centralcharge}

In this section we turn to the computation of the central charge of the SCQM. The definition of the central charge of a superconformal quantum mechanics involves some caveats related to the fact that the energy momentum tensor of a conformal quantum mechanics must vanish identically. The central charge should then be interpreted as counting the degeneracy of ground states of the system. Some proposals exist in the literature for computing this degeneracy.  In \cite{Denef:2002ru,Cordova:2014oxa,Ohta:2014ria} the number of ground states of quiver quantum mechanics with gauge group $\prod_v U(N_v)$ with the $U(N_v)$ subgroups connected by bifundamentals (so-called Kronecker quivers) was computed  by quantising the classical moduli space in the Higgs branch. The result is
\begin{equation} \label{1dcentralcharge}
\mathcal{M}=\sum_{v,w}N_v N_w -\sum_v N_v^2 +1,
\end{equation}
where $N_w$ stands for the rank of the gauge groups adjacent to a given colour group of rank $N_v$. Alternatively, it was shown in  \cite{Balasubramanian:2003kq,Balasubramanian:2009bg} that when the AdS$_2$ solution dual to a $\mathcal{N}=4$ SCQM can be obtained from an AdS$_3$ space through a null compactification (the so-called null orbifold construction) the dual SCQM is realised as a chiral half of the 2d CFT dual to the AdS$_3$ solution. The AdS$_3$ spaces considered in  \cite{Balasubramanian:2003kq,Balasubramanian:2009bg}  preserve $\mathcal{N}=(4,4)$ supersymmetries, but the result can be extrapolated to the case in which they preserve $\mathcal{N}=(0,4)$, where the SCQM simply arises upon compactification of the 2d dual (0,4) CFT. In these situations the obvious interpretation of the central charge of the SCQM is as counting the excitations of the 2d CFT, and the caveats mentioned above do not apply. Moreover, one can use the expression that allows to compute the central charge of the 2d CFT from the R-symmetry anomaly to obtain the central charge of the SCQM. This was done explicitly for the class of AdS$_2$ solutions constructed in  \cite{Lozano:2020txg}, obtained by T-dualising  a sub-class of the AdS$_3$ solutions to Type IIA supergravity with $\mathcal{N}=(0,4)$ supersymmetries constructed in  \cite{Lozano:2019emq}\footnote{It is straightforward to see that the Abelian T-duality transformation performed in  \cite{Lozano:2020txg} is equivalent to the null orbifold construction in  \cite{Balasubramanian:2003kq,Balasubramanian:2009bg}.}. The central charge computed this way was shown to agree with the holographic calculation in the holographic limit. 

Remarkably, in  \cite{Lozano:2020sae,Lozano:2021rmk,Lozano:2021fkk} other classes of AdS$_2$ solutions with $\mathcal{N}=4$ supersymmetry were constructed that do not bear any relation with AdS$_3$. Still, the expression that gives the central charge of a 2d (0,4) CFT was used to compute the central charge and it was shown to agree with the holographic result. This agreement is a remarkable result that should be investigated in more detail. It could be related to the fact that the 2d expression can be shown to agree to leading order with the 1d expression given by \eqref{1dcentralcharge}, when applied to the same type of quivers. 
In this section we show that the SCQM proposed in the previous sections provides one further example in which the central charge computed from the 2d expression is in agreement with the holographic calculation. 
Let us now show how this happens.

As mentioned, the central charge of a 2d (0,4) CFT can be computed from the R symmetry anomaly  (see for instance \cite{Putrov:2015jpa}),
\begin{equation} \label{2dcentralcharge}
c_R=3\, \text{Tr}[\gamma^3 Q_R^2],
\end{equation}
where the trace is over the Weyl fermions of the theory, $\gamma^3$ is the 2d chirality matrix and $Q_R$ is the U(1)$_R$ charge. 
Recalling the well-known facts:

\begin{itemize}
\item (0,4) vector multiplets contain two left-moving fermions with R-charge 1,
\item (0,4) twisted hypermultiplets contain two right-moving fermions with R-charge 0,
\item (0,4) hypermultiplets contain two right-moving fermions with R-charge -1,
\item (0,2) Fermi multiplets contain one left-moving fermion with R-charge 0,
\end{itemize}
and substituting in \eqref{2dcentralcharge}, one gets the well-known expression
\begin{equation}
c_R=6\, (n_{hyp}-n_{vec}),
\end{equation}
where $n_{hyp}$ stands for the number of (0,4) untwisted hypermultiplets and $n_{vec}$ for the number of (0,4) vector multiplets. If we now use this formula for our 1d quiver in Figure \ref{branebox} we find\footnote{Here we have reinstated $Q_{\text{D}6}\neq 1$.},
\begin{eqnarray}
&&n_{hyp}=\sum_{n=1}^{P-1}\sum_{m=1}^{P'-1} n^2 m (m+1) Q_{\text{D}6}^2\nonumber\\
&&n_{vec}=\sum_{n=1}^{P-1}\sum_{m=1}^{P'-1} n^2 m^2 Q_{\text{D}6}^2.
\end{eqnarray}
This gives
\begin{equation}
c_R=\frac12 P(P-1)(2P-1)P'(P'-1)Q_{\text{D}6}^2 
\end{equation}
and, to leading order in $P$ and $P'$ (large number of nodes limit),
\begin{equation}
c_R\sim P^3 P'^2 Q_{\text{D}6}^2. 
\end{equation}
In order to compare with the holographic calculation we need to compute as well $c_L$, since 
\begin{equation}
c_{hol}=\frac12 (c_L+c_R).
\end{equation}
We compute $c_L$ from
\begin{equation}
c_L=c_R+\text{Tr}\,\gamma^3.
\end{equation}
We can easily see that $\text{Tr}\gamma^3=0$ for our quiver in Figure \ref{branebox}, such that $c_L=c_R$.

Let us compute now the holographic central charge. Here we have to account for the proper normalisation of Newton's constant, as mentioned previously. The proper normalisation is
\begin{equation}
c_{hol}=\frac{3}{2^5 \pi^8}V_{int}=\frac{3}{2^5 \pi^8}\int d\vec{\theta}\,e^{-2\Phi}\sqrt{\text{det}(g_{ij})},
\end{equation}
where $g_{ij}$ is the metric of the inner space and $\vec{\theta}$ are coordinates defined over it. Using this expression and integrating $\rho$ between $[0,P\pi]$ and $\mu$ between $[0,P'\pi]$ we find 
\begin{equation}
c_{hol}=P^3P'^2 Q_{\text{D}6}^2,
\end{equation}
in perfect agreement with the field theory calculation, in the large number of nodes limit. 

\section{Conclusions}

In this paper we have constructed a new class of AdS$_2$ solutions to Type IIA supergravity with $\mathcal{N}=4$ supersymmetry realised as AdS$_2\times S^2\times S^2$ foliations over 4 intervals. These solutions arise after performing a non-Abelian T-duality transformation (with respect to a freely acting SU(2) group) on the class of solutions to Type IIB supergravity constructed in \cite{Lozano:2021fkk}. We have then focused on their defect CFT interpretation. We have shown that for a particular brane profile the resulting background asymptotes locally to a Gaiotto-Maldacena geometry, which suggests that it should be dual to a line defect CFT within the 4d $\mathcal{N}=2$ SCFT dual to this geometry. We have identified the brane set-up associated to this solution and from it we have constructed a 1d quiver quantum mechanics that, we propose, flows in the IR to the SCQM dual to the AdS$_2$ solution. We stress that the 1d quiver field theory that we have constructed is an elaborated quantum mechanics described by a D2-brane box model of the type constructed in \cite{Hanany:2018hlz}. We have shown that this quiver can be interpreted as a result of embedding the defect branes of the solution in the 4d $\mathcal{N}=2$ background theory.

 Following \cite{Lozano:2020sae}  we have given an interpretation to the massive F1-strings present in the solution in terms of baryon vertices within the 4d $\mathcal{N}=2$ SCFT, in analogy with the findings in \cite{Lozano:2020sae,Lozano:2021rmk,Lozano:2021fkk}. This is consistent with an interpretation of the AdS$_2$ solution as describing backreacted baryon vertices within the 4d $\mathcal{N}=2$ SCFT, living in a D4-NS5-D6 subsystem of the complete brane set-up. In this interpretation the SCQM arises in the very low energy limit of a system of D4-NS5-D6 branes in which one dimensional defects are introduced. The one dimensional defects consist on D4'-brane baryon vertices, connected to the D4 with F1-strings, and D2-brane baryon vertices connected to the D6 with F1-strings. In the IR the gauge symmetry on the D4-branes, that played the role of colour branes in the 4d SCFT, becomes global, turning them from colour to flavour branes. In turn, the D2-branes, stretched between the two field theory directions present in the brane set-up, become the new colour branes of the backreacted geometry. Extra NS5'-branes present in the brane set-up make this possible.
 
Our construction in this paper provides one further example of the successful applications of non-Abelian T-duality to holography \cite{Sfetsos:2010uq,Lozano:2011kb,Lozano:2012au,Itsios:2013wd,Lozano:2013oma,Macpherson:2014eza,Kelekci:2014ima,Lozano:2015bra,Lozano:2015cra,Kelekci:2016uqv,Lozano:2016kum,Lozano:2016wrs,Lozano:2017ole,Itsios:2017cew,Lozano:2018pcp,Lozano:2019ywa,Ramirez:2021tkd}. In this case it has allowed to construct a line defect CFT within a 4d $\mathcal{N}=2$ SCFT in a well-controlled string theory set-up, with a known holographic dual. As a spin-off of this construction the first holographic dual to a general brane box model has been provided\footnote{See also \cite{Faedo:2020lyw} for AdS$_3$ duals to particular D3-brane box models with one circular direction.}. Our construction provides one concrete example involving a particular 4d  $\mathcal{N}=2$ SCFT, but one would expect that a full class of AdS$_2$ solutions that asymptote in the UV to more general Gaiotto-Maldacena geometries should exist. The success of the applicability of non-Abelian T-duality in the context of holography has been the possibility to access particular solutions that have inspired the construction of more general classifications. Notable examples in the literature of this sort are the AdS solutions reported in \cite{Lozano:2012au,Macpherson:2014eza,Lozano:2015cra,Kelekci:2016uqv,Lozano:2018pcp,Lozano:2019ywa}.
Our construction in this paper could very well provide one further example that prompts an extension to a general classification of AdS$_2$ solutions asymptoting Gaiotto-Maldacena geometries in the UV.

Finally, we think it should be of interest to further explore the applications of the new geometry constructed in Appendix \ref{second} to the description of four dimensional extremal black holes with $\mathcal{N}=4$ supersymmetries.

\section*{Acknowledgements}

We would like to thank Niall Macpherson for very useful discussions.
YL and CR are partially supported by the AEI through the Spanish grant MCIU-22-PID2021-123021NBI00 and by the FICYT through the Asturian grant SV-PA-21-AYUD/2021/52177. CR is  supported by a Severo Ochoa Fellowship by the Principality of Asturias (Spain).
The work of NP is supported by the Israel Science Foundation (grant No. 741/20) and by the German Research Foundation through a German-Israeli Project Cooperation (DIP) grant ``Holography and the Swampland".

\appendix

\section{The non-Abelian T-duality transformation}\label{first}

In this appendix we give the details of the non-Abelian T-duality transformation used to construct the class of solutions presented  in section 2 (and in the following Appendix). We perform the non-Abelian T-duality with respect to a freely acting $SU(2)$ on the $S^3$ in the background \eqref{solutionsIIB}.
The sigma model describing the propagation of a string on the NS-NS sector of this background can be cast in the general form 
\begin{equation}
\label{originalL}
L=G_{\mu\nu}(X)\partial_+ X^\mu\partial_- X^\nu+G_{\mu i}(X)(\partial_+ X^\mu L_-^i+\partial_- X^\mu L^i_+)+g_{ij}(X)L^i_+ L^j_-.
\end{equation}
Here $L^i_{\pm}=-i {\rm Tr}(t^i g^{-1}\partial_\pm g)$, $g$ is an element of $SU(2)$, $t^i$ stand for the generators of $SU(2)$, and
$X^\mu$ run over AdS$_2$, $S^2$, $y$, $z$ and $r$.
Using the invariance of the sigma model  under $g\rightarrow hg$, with $h\in SU(2)$, the following non-Abelian T-dual NS-NS background can be generated (see for instance \cite{Sfetsos:1996pm})
\begin{equation}
\label{dualL}
{\tilde L}=G_{\mu\nu}\partial_+ X^\mu \partial_- X^\nu+(\partial_+ v_i +\partial_+ X^\mu G_{\mu i})M^{-1}_{ij}(\partial_- v_j-G_{\mu j}\partial_- X^\mu)\, .
\end{equation}
Here $M=g+f$, with $f$ the $SU(2)$ structure constants in the normalisation $[t^i,t^j]=i  \epsilon_{ijk}t^k$, and $g$ has been replaced by the Lagrange multipliers $v_i$, $i=1,2,3$, which take values on the Lie algebra of $SU(2)$. The Lagrange multipliers enforce the flat connection condition in the proof of equivalence between the original Lagrangian (\ref{originalL}) and its NATD (\ref{dualL})  \cite{delaOssa:1992vci}.
The dual metric and NS-NS 2-form read
\begin{eqnarray}
&&{\tilde g}_{ij}=\frac12 M^{-1}_{(ij)}\, , \qquad {\tilde B}_{ij}=\frac12 M^{-1}_{[ij]}\, ,  \qquad
{\tilde G}_{i\mu}=-\frac12 M^{-1}_{[ij]}G_{j\mu}\nonumber\\
&&{\tilde B}_{i\mu}=-\frac12 M^{-1}_{(ij)}G_{j\mu}\, , \qquad
{\tilde G}_{\mu\nu}=G_{\mu\nu}-M^{-1}_{ij}G_{\mu i}G_{\nu j}
\end{eqnarray}
where in our conventions $M_{(ij)}=M_{ij}+M_{ji}$ and $M_{[ij]}=M_{ij}-M_{ji}$.
The dilaton in turn transforms as
\begin{equation}
\Phi\rightarrow \Phi-\frac12 \log{({\rm det}M)}\, .
\end{equation}

The general procedure to generate the dual RR fields was worked out in  \cite{Sfetsos:2010uq}, and later applied in \cite{Itsios:2013wd} (see also \cite{Itsios:2012dc}) to obtain general formulas for the duals of 
$M_7\times S^3$ backgrounds with warpings  depending on the $M_7$ directions and RR fluxes consistent with this structure. 

Writing the RR field strengths of the original background as
\begin{equation}
F_p=G_p^{(0)}+G_{p-1}^a \wedge e^a+\frac12 G_{p-2}^{ab}\wedge e^a \wedge e^b+G_{p-3}^{(3)}\wedge e^1\wedge e^2\wedge e^3\, ,
\end{equation}
the ones in the NATD background\footnote{We use tildes to denote them throughout this section.}
\begin{equation}
\label{RRdual}
{\tilde F}_p={\tilde G}_p^{(0)}+{\tilde G}_{p-1}^a \wedge {\tilde e}^a+\frac12 {\tilde G}_{p-2}^{ab}\wedge {\tilde e}^a \wedge {\tilde e}^b+{\tilde G}_{p-3}^{(3)}\wedge {\tilde e}^1\wedge {\tilde e}^2\wedge {\tilde e}^3\, ,
\end{equation}
can be derived from the transformation rules
\begin{eqnarray}
&&{\tilde G}_{p}^{(0)}=\Bigl( -A_0G_p^{(3)}+A_a G_p^a \Bigr)\nonumber\\
&&{\tilde G}_{p-1}^a=\Bigl( -\frac{A_0}{2} \epsilon^{abc}G_{p-1}^{bc}+A_b G_{p-1}^{ab}+A_a G_{p-1}^{(0)}\Bigr) \nonumber\\
&&{\tilde G}_{p-2}^{ab}=\Bigl[ \epsilon^{abc}\Bigl( A_c G_{p-2}^{(3)}+A_0 G_{p-2}^c \Bigr)-(A_a G_{p-2}^b-A_b G_{p-2}^a)\Bigr] \nonumber\\
&&{\tilde G}_{p-3}^{(3)}=\Bigl(\frac{A_a}{2}\epsilon^{abc}G_{p-3}^{bc}+A_0 G_{p-3}^{(0)}\Bigr)\, .
\end{eqnarray}
Here $a,b,c$ run over the directions of the $S^3$ on which the dualisation is performed, which here is taken to be squashed for generality, $e^a$ are the frames on these directions and the coefficients $A_0$ and $A_a$ are given by
\begin{equation}
A_0=e^{B_1+B_2+B_3}\, , \qquad A_a = v_a e^{B_a}\, ,
\end{equation}
where $A_a$ and $B_a$ are functions of the spectator coordinates, $v_i$ are the Lagrange multipliers living in the Lie algebra of $SU(2)$, and we have assumed an original geometry of the form
\beq
ds^2_{str} = g_{\mu\nu}dx^{\mu}dx^{\nu}+ e^{2B_a}(\omega^a +A^a)^2,
\eeq
where $\omega^a$ are left invariant 1-forms and $A^a = A^a_{~\mu}dx^{\mu}$ are fibration terms.

\section{A black hole geometry through non-Abelian T-duality}\label{second}

In this Appendix we present the non-Abelian T-dual (with respect to a freely acting SU(2), as reviewed in the previous Appendix) acting on the brane solution that underlies the AdS$_2$ solutions \eqref{solutionsIIB}, given by equation (3.8) in \cite{Lozano:2021fkk}. One can check that the solutions  \eqref{NATDyz} arise in the near horizon limit of this solution. However we have not succeeded in given a concrete interpretation in terms of a brane intersection, that would be underlying the near-horizon geometries  \eqref{NATDyz}. As we have explained in the main text this is a common feature for AdS solutions constructed through non-Abelian T-duality. Still, the solution presented in this section  should have an interesting interpretation as describing 4d $\mathcal{N}=4$ extremal black holes. 

The solution is given by:

\begin{equation}
	\label{sol:D1-F1-D5-NS5-D3_NATD}
	\begin{split}
		d s_{10}^2 &= H_{\text{D}3}^{-1/2}  \Big[- H_{\text{D}1}^{-1/2} H_{\text{F}1}^{-1} H_{\text{D}5}^{-1/2}dt^2 +H_{\text{D}1}^{1/2}  H_{\text{D}5}^{1/2}H_{\mathrm{NS}5}\left(d\zeta^2+\zeta^2ds^2_{S^2} \right) +\\
		&+ 4r^{-2} H_{\text{D}1}^{-1/2}  H_{\text{D}5}^{1/2}(d\rho^2+  H \rho^2 ds^2_{\tilde S^2}) \Big] + H_{\text{D}3}^{1/2} \Big[H_{\text{D}1}^{-1/2}  H_{\text{D}5}^{-1/2}H_{\mathrm{NS}5}dy^2+\\
		& + H_{\text{D}1}^{1/2} H_{\text{F}1}^{-1} H_{\text{D}5}^{1/2}dz^2 + H_{\text{D}1}^{1/2}  H_{\text{D}5}^{-1/2}\,dr^2\Big]\,,\\
		e^{\Phi}&= (r/2)^{-3} H^{1/2} H_{\text{D}1}^{-1/4}  H_{\text{F}1}^{-1/2} H_{\text{D}5}^{1/4} H_{\mathrm{NS}5}^{1/2} H_{\text{D}3}^{-3/4}\,,\\
		B_{2} &= z\partial_\zeta H_{\text{F}1}^{-1}\, dt\wedge d\zeta - y \zeta^2 \partial_\zeta H_{\mathrm{NS}5} \, \text{vol}_{S^2}+\frac{16 H_{\text{D}5} \rho^3}{16 \rho^2 H_{\text{D}5} + H_{\text{D}1} H_{\text{D}3} r^4} \, \text{vol}_{\tilde{S}^2}\,,\\
		F_{2} &= -8^{-1} r^3 \left[\left( H_{\text{F}1}H_{\text{D}5}^{-1}\partial_z H_{\text{D}3} \, dy - H_{\text{D}1} H_{\mathrm{NS}5}^{-1}\partial_y H_{\text{D}3} \, dz\right) \wedge dr - \partial_r H_{\text{D}3} \, dy \wedge dz\right]\,, \\
		F_{4} &= - 8^{-1} r^3  H_{\text{D}3}(\partial_\zeta H_{\text{D}5}^{-1} dt \wedge d\zeta \wedge dy + \zeta^2 \partial_\zeta H_{\text{D}1} \text{vol}_{S^2} \wedge dz) \wedge dr +\\
		&+ \rho(\partial_\zeta H_{\text{D}1}^{-1} dt \wedge d\zeta \wedge dy + \zeta^2 \partial_\zeta H_{\text{D}5} \text{vol}_{S^2} \wedge dz) \wedge d\rho+\\
		&+ \frac{16 H_{\text{D}5} \rho^3}{16 \rho^2 H_{\text{D}5} +  H_{\text{D}1} H_{\text{D}3} r^4} \, \text{vol}_{\tilde{S}^2}\wedge F_{2}  \,,\\
		F_{6} &= d[H_{\text{D}5}H_{\mathrm{NS}5}H_{\text{D}3}^{-1} \rho \zeta^2 \,dt\wedge d\zeta\wedge \text{vol}_{ S^2}\wedge d\rho]+\\
		& - \rho^2 H [2r^{-2} \rho H_{\text{D}1}^{-1} H_{\text{D}5} (\partial_\zeta H_{\text{D}5}^{-1} dt \wedge d\zeta \wedge dy + \zeta^2 \partial_\zeta H_{\text{D}1} \text{vol}_{S^2} \wedge dz) \wedge dr + \\
		&+ (\partial_\zeta H_{\text{D}1}^{-1} dt \wedge d\zeta \wedge dy +\zeta^2 \partial_\zeta H_{\text{D}5} \text{vol}_{S^2} \wedge dz) \wedge d\rho] \wedge \text{vol}_{\tilde{S}^2}\,,
	\end{split}
\end{equation} 
where we have defined
\begin{equation}
	H = \frac{H_{\text{D}1} H_{\text{D}3} r^4}{16 \rho^2 H_{\text{D}5} + H_{\text{D}1} H_{\text{D}3} r^4}\,.
\end{equation}

\end{document}